\documentclass[12pt]{article}
\usepackage[utf8]{inputenc}
\usepackage{natbib}
\usepackage{mathtools, amsthm}
\usepackage{enumerate}
\usepackage{array}
\usepackage{times}
\usepackage{amsmath}
\usepackage{latexsym}
\usepackage{amssymb}
\usepackage{graphics}
\usepackage{graphicx}
\usepackage{accents}
\usepackage{color}
\usepackage{mathrsfs}
\usepackage[normalem]{ulem}
\usepackage{url}
\usepackage{wrapfig}
\usepackage[font=scriptsize,labelfont=bf]{caption}
\usepackage{multirow}
\usepackage{algorithm} 
\usepackage{algpseudocode} 
\usepackage{natbib}
\usepackage{subfig}
\usepackage{setspace} 
\usepackage{bm}

\def\bs{\boldsymbol}

\newcommand{\smallnum}[1]{\multicolumn{1}{c}{\small #1}}

\definecolor{red}{rgb}{1,0,0}
\definecolor{green}{rgb}{0,1,0}
\definecolor{blue}{rgb}{0,0,1}

\definecolor{purple}{RGB}{180,0,180}

\def\bs{\boldsymbol}

\newcommand{\Real}{\mathbb{R}} 
\newcommand{\Tra}{^{\sf T}} 

\newcommand{\V}[1]{{\bm{\mathbf{\MakeLowercase{#1}}}}} 
\newcommand{\Vn}[2]{\V{#1}^{(#2)}} 

\newcommand{\M}[1]{{\bm{\mathbf{\MakeUppercase{#1}}}}} 

\newcommand{\prox}{\mathop{\rm prox}\nolimits}

\newcommand{\df}{\mathop{\rm df}\nolimits}

\newcommand{\amp}{\mathop{\:\:\,}\nolimits}

\usepackage{comment}
\usepackage{xspace}
\definecolor{blue}{rgb}{0.2,0.5,0.7}
\definecolor{green}{rgb}{0.3,0.68,0.29}
\definecolor{purple}{rgb}{0.6,0.31,0.64}

\definecolor{sy}{RGB}{229,114,0}

\newtheorem{theorem}{Theorem}
\newtheorem{lemma}{Lemma}
\newtheorem{proposition}{Proposition}
\newcommand{\blind}{1}

 \makeatletter
\newcommand*{\addFileDependency}[1]{
  \typeout{(#1)}
  \@addtofilelist{#1}
  \IfFileExists{#1}{}{\typeout{No file #1.}}
}
\makeatother

\begin{document}
\def\spacingset#1{\renewcommand{\baselinestretch}%
{#1}\small\normalsize} \spacingset{1}

\if1\blind
{ 
  \title{\bf Robust Spatiotemporal Epidemic Modeling with Integrated Adaptive Outlier Detection}
  \author{Haoming Shi\\
    Department of Statistics, Rice University\\
    Shan Yu \\
    Department of Statistics, University of Virginia \\
    Eric C. Chi \\
    School of Statistics, University of Minnesota}
  \date{}
  \maketitle
} \fi

\if0\blind
{
\title{\bf Robust Spatiotemporal Epidemic Modeling with Integrated Adaptive Outlier Detection}
  \bigskip
  \bigskip
  \bigskip
  \medskip
  \date{}
  \maketitle
} \fi

\bigskip
\begin{abstract}
In epidemic modeling, outliers can distort parameter estimation and ultimately lead to misguided public health decisions. Although there are existing robust methods that can mitigate this distortion, the ability to simultaneously detect outliers is equally vital for identifying potential disease hotspots. In this work, we introduce a robust spatiotemporal generalized additive model (\texttt{RST-GAM}) to address this need. We accomplish this with a mean-shift parameter to quantify and adjust for the effects of outliers and rely on adaptive Lasso regularization to model the sparsity of outlying observations. We use univariate polynomial splines and bivariate penalized splines over triangulations to estimate the functional forms and a data-thinning approach for data-adaptive weight construction. We derive a scalable proximal algorithm to estimate model parameters by minimizing a convex negative log-quasi-likelihood function. Our algorithm uses adaptive step-sizes to ensure global convergence of the resulting iterate sequence. We establish error bounds and selection consistency for the estimated parameters and demonstrate our model's effectiveness  through numerical studies under various outlier scenarios. Finally, we demonstrate the practical utility of \texttt{RST-GAM} by analyzing county-level COVID-19 infection data in the United States, highlighting its potential to inform public health decision-making.
\end{abstract}

\noindent%
\vfill

\newpage

\spacingset{1.8}

\section{Introduction}
Spatiotemporal epidemic modeling has received significant attention in recent years in the wake of the devastating COVID-19 pandemic \citep{giuliani2020modelling, nikparvar2021spatio, tsori2022spatio}. Understanding spatial transmission patterns and the impact of covariates on infection rates is essential to prevention strategies and resource allocation. Regression models provide a way to gain this understanding by identifying actionable associations between a response variable and a set of explanatory variables. The scale and complexity of spatiotemporal data, however, present unique challenges. The relationship between infection counts and covariates are often nonlinear and spatially heterogeneous \citep{schuster2022modeling}. A parametric regression model may fail to capture these intricate dynamics due to rigid over-simplification, leading to potential biases and loss of critical insights. This limitation motivates nonparametric alternatives \citep{smirnova2019forecasting, Yu2020, wang2022}. In particular, generalized additive models (GAMs) are widely used since they flexibly model the mean of the response variable as an additive combination of smooth functions of the predictors through a specified link function \citep{hastie1987generalized}. For example, \cite{de2021using} combine a GAM and Markov-Switching model to detect shifts in disease trends.  \cite{wang2022} integrate a GAM into a mathematical epidemiology framework to capture disease spread dynamics. \cite{eckardt2024generalized} extend a GAM to compositional data analysis. Nonparametric epidemic models are significantly less prone to bias than parametric ones, enabling deeper exploration of disease dynamics and providing actionable insights.

The effectiveness of GAMs, however, is confounded by the presence of outliers. Outliers are pervasive in statistical analysis, with routine datasets typically containing about 1–10\% outliers \citep{hampel1986robust}. They may contaminate the response or explanatory variables, potentially compromising the model estimation and inference \citep{chatterjee1988sensitivity}. In the context of epidemiology, outliers, often observed as upward spikes in infection counts, are particularly prevalent \citep{meirom2014localized}. For instance, areas with generally low infection counts may occasionally include specific regions experiencing sudden surges in infections compared to their neighbors. These spikes can arise due to reporting errors, localized outbreaks, or other irregularities \citep{roukema2020anomalies}. Unaddressed outliers bias parameter estimates, jeopardizing inference about disease transmission patterns. In this work, we address the challenge that created by spikes in infection counts when using GAMs for spatiotemporal epidemic analysis.

In the statistics literature, there are two primary directions for handling outliers. The first direction mitigates the influence of outliers by modifying the objective function, e.g., M-Estimators \citep{huber1964robust} and quantile regression \citep{koenker1978regression}. These approaches are not primarily concerned with identifying outlying observations. By contrast, the second direction focuses on outlier detection, seeking to identify and potentially remove observations that deviate significantly from the typical sample using methods such as distance-based \citep{knorr1998}, density-based \citep{breunig2000}, and clustering techniques \citep{he2003}. While both directions have their merits, addressing outliers in epidemic modeling requires a unified framework. On the one hand, estimators should remain unbiased and insensitive to the presence of spikes. On the other hand, the model should identify outlying observations that correspond to transmission hotspots, since this is actionable information for public health interventions. Hence, we seek a spatiotemporal epidemic model that integrates robustness and outlier detection to address these dual needs.

Our solution is to incorporate sparse mean-shift parameters into the regression model \citep{CandesLiMaWright2011, she2011outlier}. Doing so enables identifying outlying responses and estimating the primary parameters of interest robustly. The core idea is to assign each observation with a mean-shift parameter that is penalized to incentivize sparsity in it. These additional parameters capture outlying deviations of responses from their expected values, thereby improving the reliability of model estimation. This approach has been extensively studied in regression settings and has proven effective for achieving both robustness and outlier detection \citep{kong2018fully, liao2020outlier}. Incorporating mean-shift parameters into GAMs within the spatiotemporal epidemic framework presents two challenges, however. First, the strategy's effectiveness heavily relies on the choice of regularization.  \cite{she2011outlier} show that a Lasso penalty, while promoting sparsity, fails to guarantee estimation and selection consistency. Design of the penalty function is critical. Second, estimation is computationally challenging due to the size of pandemic datasets. Implementing a scalable penalized GAM framework  necessitates algorithmic innovation.

To address these challenges, we propose a robust spatiotemporal generalized additive model (\texttt{RST-GAM}) by combinin the mean-shift framework with the adaptive Lasso. This approach addresses the dual objectives of robustness and outlier detection in the spatiotemporal epidemic context. Since our focus is on spikes in infection counts, we adopt terminology from the optimization field \citep{dantzig1951linear} and refer to these mean-shift parameters as slack variables to reflect their positivity. The main contributions of this work are three-fold:
\begin{itemize}
    \item We develop a robust class of GAMs tailored for spatiotemporal epidemic data by incorporating slack variables into the regression framework. This integration enables simultaneous robust estimation and outlier detection. To improve the precision of outlier detection, we employ adaptive Lasso regularization, which ensures sparsity and selection consistency.
    \item We introduce a novel method for constructing adaptive weights based on the data-thinning technique. 
    \item Finally, we design a proximal gradient descent (PGD) algorithm with adaptive step-sizes to solve the resulting regularized optimization problem. This algorithm requires no additional computational effort beyond gradient evaluations and guarantees convergence for convex and locally Lipschitz problems like the one defined by the \texttt{RST-GAM} model fitting task.
\end{itemize}

The rest of this paper is organized as follows. In Section~\ref{sec:method}, we describe the proposed  framework, including spline approximations for component functions and the weight construction for adaptive Lasso. In Section~\ref{sec:theoretical_results}, we present theoretical guarantees for estimation and selection consistency of \texttt{RST-GAM}. In Sections~\ref{sec:algorithm} and \ref{sec:model_selection}, we discuss the optimization algorithm and tuning parameter selection. In Section \ref{sec:simulation}, we compare the performance of \texttt{RST-GAM} against a standard GAM approach through numerical studies, highlighting its robustness and accuracy in outlier detection. In Section~\ref{sec:Real_Covid}, we demonstrate  \texttt{RST-GAM}'s utility on COVID-19 infection data. In Section~\ref{sec:conclusion}, we conclude and discuss potential extensions of our work. Proofs and technical details are given in the Supplementary Materials.

\section{The \texttt{RST-GAM} Method}
\label{sec:method}
At each time $t \in [T]$ and location $i \in [n]$, where $[n] = \{1,2,\dots, n\}$, we observe two variables: $Y_{it}$, the number of new infections and $\M{X}_{it}=(X_{it1}, \ldots, X_{itp}){\Tra} \in \Real^p$, a vector of explanatory variables. Let $\V{u}_i=(u_{i1}, u_{i2}) \in \Omega \subset \Real^2$ be the spatial coordinates of the $i$th location; assume the region $\Omega$ is bounded. Let $\mu_{it}(\V{u}_i, \V{x}_{it}) = \mathbb{E}\left[ Y_{it}|\M{U} = \V{u}_i, \mathbf{X} = \V{x}_{it} \right]$ be the conditional mean of $Y_{it}$. We model the conditional mean $\mu_{it}(\V{u}_i, \V{x}_{it})$ using a GAM \citep{Yu2020}, i.e., 
\begin{eqnarray}
\label{eq:mean_clean}
g(\mu_{it}) & = & \beta_{t}(\V{u}_i) + \sum_{k=1}^{p}\alpha_{kt}(X_{itk}),
\end{eqnarray}
where (i) $g$ is a link function, (ii) $\beta_{t} : \Omega \to \Real$ is a bivariate smooth function of the spatial coordinates $\V{u}$, representing a baseline spatial transmission rate for each location at time $t$, and (iii) $\alpha_{kt} \colon \Real \to \Real$ for $k \in [p]$ are univariate functions that model the effect of covariate $X_k$ on $g(\mu_{it})$, where $X_k$ denotes a random variable for covariate $k \in [p]$. We assume $\mathbb{E}[\alpha_{kt}(X_k)] = 0$ to ensure that the GAM is identifiable \citep{hastie1986generalized}. 

To identify  outliers and estimate $\beta_{t}$ and $\alpha_{kt}$ robustly, we introduce a nonnegative slack variable $\xi_{it}$ as a mean shift component in \eqref{eq:mean_clean},
\begin{eqnarray}
g(\mu_{it}) & = & \beta_{t}(\V{u}_i) + \sum_{k=1}^{p}\alpha_{kt}(X_{itk}) + \xi_{it}. 
\label{eq:mean_slack}
\end{eqnarray}
If $\xi_{it}$ is positive, the observation $Y_{it}$ is an outlier. Then the conditional mean $\mu_{it}$ now also depends on $\xi_{it}$ as $\mu_{it}(\V{u}_i, \mathbf{x}_{it}, \xi_{it}) = \mathbb{E}\left[ Y_{it}|\M{U} = \V{u}_i, \mathbf{X} = \V{x}_{it}, \mathbf{\Xi} = \xi_{it} \right]$. 


\subsection{Robust Penalized Quasi-Likelihood Estimation}

We assume that the conditional variance of $Y$ given $\V{u}$, $\V{x}$, and $\V{\xi}$ can be expressed as 
\begin{eqnarray*}
\textrm{Var}(Y|\M{U}  = \V{u},\mathbf{X} = \V{x}, \mathbf{\Xi} = \V{\xi}) & = & \sigma^2V(\mu(\V{u}, \mathbf{x}, \V{\xi}))
\end{eqnarray*}
where $V$ is a known positive function and $\sigma$ is a positive dispersion parameter. We estimate the conditional mean $\mu_{it}$ by maximizing a penalized quasi-likelihood function $L(\mu,y)$ that  satisfies $\nabla_\mu L(\mu,y) = \frac{y-\mu}{\sigma^2V(\mu)}$. In this work, we model the number of new cases with a Poisson distribution. Consequently, $V(\mu) = \mu$ and $g(\mu) = \log(\mu)$. 
We estimate $\beta_t, \alpha_{kt},$ and $\V{\xi}_t$ with the solution to the optimization problem
\begin{equation}
    \label{eq:Model_Estimation_1}
    \underset{\beta_t, \,\alpha_{kt}, \,\V{\xi}_t \geq \V{0}}{\text{minimize}} -\sum_{i=1}^{n}\sum_{s=t-t_0}^{t} L\left (\exp\left\{\beta_{t}(\V{u}_i)+ \sum_{k=1}^{p}\alpha_{kt}(X_{isk}) + \xi_{it}\right\}, Y_{is}\right)+ \frac{\lambda_{0}}{2}\mathcal{E}(\beta_{t})  + \lambda_1 \|\V{\xi}_t\|_1,
\end{equation}
where $t_0+1$ is the size of a moving window that captures the temporal dynamics. The penalty functions $\mathcal{E}(\beta)$ and $\lVert \V{\xi} \rVert_1$ impose structure in the bivariate components $\beta_t$ and the slack variables $\xi_{it}$. The hyperparameters $\lambda_0$ and $\lambda_1$ tradeoff model fit and model complexity. 

We elaborate on our choice of the penalty functions. The roughness penalty  $\mathcal{E}(\beta_t)$ on $\beta_t$ encourages spatial smoothness.
\begin{eqnarray}
	\mathcal{E}(\beta)
	& = & \int_{\Omega} \left\{(\nabla_{u_{1}}^{2}\beta)^2+2(\nabla_{u_{1}}\nabla_{u_{2}}\beta)^2+(\nabla_{u_{2}}^{2}\beta)^2\right\}\mathrm{d}u_{1}\mathrm{d}u_{2},
\label{EQ:energyfun}
\end{eqnarray}
where $\nabla_{u_{j}}^{q}\beta(\V{u})$ is the $q$th order directional derivative at location $\V{u}=(u_1,u_2)$ in the direction of $u_{j}$, $j=1,2$. The Lasso penalty on $\V{\xi}_t = \{\xi_{1t}, \ldots, \xi_{nt}\} \in \Real^{n}$ encourages sparsity in the slack variable to model outlying count observations as rare. We assume these outliers' effects depend only on location $\boldsymbol{u}_i$ and are uniform over an observation period $t_0 + 1$. We also impose a non-negativity constraint, i.e.,  $\xi_{it} \in \Real_+$, as this work focuses on the effects of spikes in observed infection counts. We can relax this constraint if we wish to model outlying low counts.

Our estimation procedure consists of two main components to simultaneously estimate $\beta_t$ and $\alpha_{kt}$ robustly and identify outliers through $\xi_{it}$. First, the nonparametric nature of the univariate $\alpha_{kt}$ and bivariate $\beta_{t}$ functions make the estimation challenging, especially when the spatial domain $\Omega$ contains holes and irregular shapes with complicated boundaries. We address this challenge by approximating each $\alpha_{kt}$ with univariate polynomial splines and $\beta_{t}$ with the bivariate penalized splines over triangulation (\texttt{BPST}), which has been shown to be effective for data distributed on complex domain \citep{BPSTMethod, Yu2020}. 

Second, outliers are identified by positive $\xi_{it}$. Outliers are rare. Consequently, we seek a sparse estimate of $\xi_{it}$. A first thought is to apply a Lasso penalty on $\xi_{it}$ \cite{tibshirani1996regression}. The Lasso, however,  suffers from shrinkage bias and variable selection inconsistency \citep{zhao2006model}. Consequently, we employ the adaptive Lasso, which reduces bias and achieves selection consistency via data-dependent weights \citep{zou2006, huang2008adaptive}. 

\subsection{Modeling the Univariate and Bivariate Functions}
\label{subsec:spline_approximation}
\noindent We first provide an approximation for univariate components. Suppose that covariates $X_k$ lie within a compact interval $[a_k, b_k]$ for $k \in [p]$. We approximate the univariate functions $\alpha_{kt}(X_k)$ from the space of polynomial splines with order $\varrho + 1$, denoting as $\mathcal{U}_k = \mathcal{U}_k^{\varrho}([a_k, b_k])$. To satisfy the constraint $\mathbb{E}[\alpha_{kt}(X_k)] = 0$, we consider the centered spline space $\mathcal{U}_k^0 = \{ \mathcal{A} \in \mathcal{U}_k,\, \mathbb{E}[\mathcal{A}(X_k)] = 0 \}$ \citep{Xue2006, Wang2007, xue2010}. Then, let $\mathcal{J}$ be the index set of the basis functions, and construct the spline basis $\{ \mathcal{A}_{kj}(X_k), \, j \in \mathcal{J} \}$ in $\mathcal{U}_k^0$ such that $\mathbb{E}[\mathcal{A}_{kj}(X_k)] = 0$ and $\mathbb{E}[\mathcal{A}_{kj}(X_k)^2] = 1$ \citep{Yu2020}. We then approximate the univariate components $\alpha_{kt}(X_k)$ by $\hat{\alpha}_{kt}(X_k)$ by expanding them in the basis functions
\begin{eqnarray}
    \hat{\alpha}_{kt}(X_k) &=& \sum_{j \in \mathcal{J}} \theta_{ktj} \mathcal{A}_{kj}(X_k) \amp = \amp \M{A}_k(X_k){\Tra} \V{\theta}_{kt}
    \label{eq:univariate_basis}
\end{eqnarray}
where $\M{A}_k(X_k) = \{ \mathcal{A}_{kj}(X_k), \, j \in \mathcal{J} \}{\Tra}$ is the vector of basis functions evaluated at $X_k$ and $\V{\theta}_{kt} = \{\theta_{ktj}, \, j \in \mathcal{J} \}{\Tra}$ is the corresponding coefficient vector to be estimated. We denote the vector of all univariate basis coefficients at time $t$ by  $\V{\theta}_t = (\V{\theta}_{1t}{\Tra}, \ldots, \V{\theta}_{pt}{\Tra}){\Tra}$.

We use \texttt{BPST} to estimate the bivariate components $\beta_t$. \texttt{BPST} is particularly effective for spatial data distributed on complex domains with irregular shapes or containing holes \citep{BPSTMethod}. For a given spatial domain $\Omega$, a {\it triangulation} of $\Omega$ is a collection of triangles $\triangle = \{\tau_1 , \dots, \tau_N \}$ that partitions $\Omega$, i.e., $\Omega = \bigcup_{i=1}^N \tau_i$, such that any two triangles share at most one common vertex or edge \citep{lai2007}. Let $(b_1, b_2, b_3)$ be the barycentric coordinates of an arbitrary point $\V{u} \in \Real^2$ relative to a triangle $\tau \in \triangle$. The Bernstein basis polynomials of degree $d \geq 1$ on $\tau$ are given as $B_{\tau,d}^{ijk}(\V{u}) = \frac{d!}{i!\, j!\, k!}\, b_1^i\, b_2^j\, b_3^k$ with constraint $i + j + k = d$. 

Denote $\mathbb{P}_d(\tau)$ as the space of all polynomials of degree up to $d$ on $\tau$, and let $\mathbb{C}^r(\Omega)$ be the set of $r$th continuously differentiable functions over $\Omega$. We define the spline space of degree $d$ and smoothness $r$ over triangulation $\triangle$ as $\mathbb{S}_d^r(\triangle) = \left\{ \zeta \in \mathbb{C}^r(\Omega) \,:\, \zeta|_{\tau} \in \mathbb{P}_d(\tau),\ \forall\, \tau \in \triangle \right\}$ and construct the corresponding set of bivariate Bernstein basis polynomials as $\{B_{m}(\V{u})\}_{m \in \mathcal{M}}$, where $\mathcal{M}$ is the basis index set. We generate these Bernstein basis polynomials via R package ``BPST'' \citep{BPST}. We then approximate the bivariate component $\beta_{t}$ in (\ref{eq:Model_Estimation_1}) with a function $\hat \beta_{t} \in \mathbb{S}_d^r(\triangle)$ through basis expansion as 
\begin{eqnarray}
    \hat \beta_{t}(\V{u}) &=& \sum_{m \in \mathcal{M}} B_{m}(\V{u})\gamma_{mt} = \mathbf{B}(\V{u}){\Tra}\V{\gamma}_t
    \label{eq:bivariate_basis}
\end{eqnarray}
where $\mathbf{B}(\V{u}) = \{B_m(\V{u}),m \in \mathcal{M}\}{\Tra}$ is the vector of basis functions evaluated at $\V{u}$, and $\V{\gamma}_t = \{ \gamma_{mt}, m \in \mathcal{M}\}{\Tra}$ is the vector of coefficients to be estimated. Substituting  approximations (\ref{eq:univariate_basis}) and (\ref{eq:bivariate_basis}) into the objective function (\ref{eq:Model_Estimation_1}) gives us the following optimization problem
\begin{align}
    \underset{\V{\gamma}_t, \, \V{\theta}_{kt}, \,\V{\xi}_t \geq \V{0}}{\text{minimize}} -\sum_{i=1}^{n}\sum_{s=t-t_0}^{t} & L\left[\exp \left\{ \mathbf{B}(\V{u}_{i}){\Tra}\V{\gamma}_t +\sum_{k=1}^{p}\M{A}_k(X_{isk}){\Tra}\V{\theta}_{kt} + \xi_{it}\right\}, Y_{is}\right] \nonumber \\&+\frac{\lambda_{0}}{2}\V{\gamma}_t{\Tra}\M{P}\V{\gamma}_t +
		\lambda_1 \|\V{\xi}_t\|_1 \quad \text{subject to} \; \mathbf{\Psi}\V{\gamma}_t =\V{0}.
  \label{eq:Model_Estimation_2}
\end{align}  
The roughness penalty function $\mathcal{E}(\beta_t)$ is approximated by $\V{\gamma}_t{\Tra}\M{P}\V{\gamma}_t$, with $\M{P}$ as a block diagonal matrix. The equality constraint $\M{\Psi}\V{\gamma}_t = \V{0}$ ensures that 
polynomials join smoothly on adjacent triangles. The matrix $\M{\Psi}$ encodes these smoothness conditions \citep{Yu2020}.

We simplify the problem by eliminating the linear constraint $\M{\Psi}\V{\gamma}_t = \V{0}$. We achieve this by reparameterizing the coefficient vector $\V{\gamma}_t$. Denote the QR decomposition of $\M{\Psi}{\Tra}$ with rank $r_{\Psi}$ as $\M{\Psi}{\Tra} = \M{Q}\M{R} = \begin{pmatrix} \M{Q}_1 \, \M{Q}_2 \end{pmatrix} \begin{pmatrix} \M{R}_1 \\ \M{R}_2 \end{pmatrix}$, where $\M{Q}$ is orthogonal and $\M{R}$ is upper triangular. The matrix $\M{Q}_1$ consists of the first $r_{\Psi}$ columns of $\M{Q}$, and $\M{R}_2$ is a matrix of zeros. Then, the reparametrization of $\V{\gamma}_t = \M{Q}_2 \V{\gamma}_t^*$ for some $\V{\gamma}_t^*$ will guarantee $\M{\Psi}\V{\gamma}_t =\V{0}$. Therefore,  problem (\ref{eq:Model_Estimation_2}) is equivalent to the simpler problem
\begin{align}
        \underset{\V{\gamma}_t, \, \V{\theta}_{kt}, \,\V{\xi}_t \geq \V{0}}{\text{minimize}} -\sum_{i=1}^{n}\sum_{s=t-t_0}^{t} & L\left[\exp\left\{\mathbf{B} (\V{u}_i){\Tra}\M{Q}_2\V{\gamma}^*_t+ \sum_{k=1}^{p}\M{A}_k(X_{isk}){\Tra}\V{\theta}_{kt} + \xi_{it}\right\}, Y_{is}\right] \nonumber \\
    & + \frac{\lambda_{0}}{2} (\V{\gamma}^*_t){\Tra}\M{Q}_2{\Tra}\M{P}\M{Q}_2\V{\gamma}^*_t + \lambda_1 \|\V{\xi}_t\|_1.
\label{eq:Model_Estimation_3}
\end{align}
The estimators of univariate and bivariate components $\alpha_{kt}$ and $\beta_t$ are then given as $\hat{\alpha}_{kt}(X_k) = {\M{A}_k(X_k)}{\Tra} \hat{\V{\theta}}_{kt}$ and $\hat{\beta}_t(\V{u}) = {\mathbf{B}(\V{u})}{\Tra} \M{Q}_2 \hat{\V{\gamma}}_t^*$, where $\hat{\V{\theta}}_{kt}$ and $\hat{\V{\gamma}}_t^*$ are the solutions of  \eqref{eq:Model_Estimation_3}.

\subsection{Modeling the Outliers}

We detect outliers via non-zero estimates of $\xi_{it}$. On the one hand, outliers are influential when they are present. On the other hand, outliers are infrequent. Consequently, we employ the adaptive Lasso to ensure selection consistency of outlying locations.  The adaptive Lasso's performance hinges on well constructed  weights, which are typically derived from Lasso or Ridge estimates. A standard way to constructing weights is to average coefficient estimates across subsamples via sample splitting or bootstrapping to improve stability \citep{wang2011randomlasso}. However, such approaches would omit some locations in each fold. This is problematic since
we require estimates of $\xi_{it}$ at each location $\V{u}_i$. 
Instead, we apply a data-thinning approach \citep{neufelddatathinning} to split infection counts $Y$ into multiple equal-sized folds $Y^{(q)}$ and compute  adaptive weights by averaging the q-fold estimates $\V{\xi}_{t}^{(q)}$ estimates. Consequently, our estimation problem (\ref{eq:Model_Estimation_3}) becomes: 
\begin{align}
        \underset{\V{\gamma}_t, \, \V{\theta}_{kt}, \,\V{\xi}_t \geq \V{0}}{\text{minimize}} -\sum_{i=1}^{n}\sum_{s=t-t_0}^{t} & L\left[\exp\left\{\mathbf{B} (\V{u}_i){\Tra}\M{Q}_2\boldsymbol{\gamma}^*_t+ \sum_{k=1}^{p}\M{A}_k(X_{isk}){\Tra}\boldsymbol{\theta}_{kt} + \xi_{it}\right\}, Y_{is}\right] \nonumber \\
    & + \frac{\lambda_{0}}{2} (\boldsymbol{\gamma}^*_t){\Tra}\M{Q}_2{\Tra}\M{P}\M{Q}_2\boldsymbol{\gamma}^*_t + \lambda_1 \sum_{i=1}^n w_{it} \lvert\xi_{it}\rvert
\label{eq:Model_Estimation_4}
\end{align}
where $\V{w}_t = (w_{1t}, \dots, w_{nt}){\Tra}$ is the weights vector. When $\V{w}_t = \V{1}{\Tra}$, problem \eqref{eq:Model_Estimation_4} reduces to the standard Lasso formulation in \eqref{eq:Model_Estimation_3}. We detail how we use data-thinning to compute data-dependent weights in Section 1 of the Supplementary Materials

\section{Properties of the \texttt{RST-GAM} Estimator}
\label{sec:theoretical_results}
In this section, we discuss the theoretical properties of the \texttt{RST-GAM}  estimator. We first introduce an oracle estimator. Without loss of generality, assume the first $h$ elements of $\bs{\xi}_t$ are nonzero, and the rest are zero. 
More specifically, define a non-zero index set $\mathcal{N}_t = \{1, \ldots, h\}$ and zero index set $\mathcal{N}_t^c = \{h+1, \ldots, n\}$. The oracle estimator $\left \{ \widehat{\bs{\gamma}}^{\ast o}_t, \widehat{\bs{\theta}}^o_t, \widehat{\bs{\xi}}^o_t\right\}$ is the maximizer of the following objective function
\begin{align*}
         & \sum_{i \in \mathcal{N}_t} \sum_{s=t-t_0}^{t} L\left[g^{-1}\left\{\mathbf{B} (\V{u}_i){\Tra}\M{Q}_2\V{\gamma}^{\ast o}_t+ \sum_{k=1}^{p+1}\M{A}_k(X_{isk}){\Tra}\V{\theta}_{kt}^o + \xi_{it}^o\right\}, Y_{is}\right] \\
         & + \sum_{i \in \mathcal{N}_t^c}\sum_{s=t-t_0}^{t} L\left[g^{-1}\left\{\mathbf{B} (\V{u}_i){\Tra}\M{Q}_2\V{\gamma}^{\ast o}_t+ \sum_{k=1}^{p+1}\M{A}_k(X_{isk}){\Tra}\V{\theta}_{kt}^o \right\}, Y_{is}\right]-\frac{\lambda_{0}}{2} (\V{\gamma}^{\ast o}_t){\Tra}\M{Q}_2{\Tra}\M{P}\M{Q}_2\V{\gamma}^{\ast o}_t,
\end{align*}
subject to $\xi_{it}^o \in \Real_+$ and $\widehat{\beta}^{o}(\bs{u}) = \widetilde{\mathbf{B}}(\bs{u})\widehat{\bs{\gamma}}^{\ast o}$ and $\widehat{\alpha}_k^{o}(x) = \M{A}_k(x){\Tra}\widehat{\bs{\theta}}_k^{o}$.
The following lemma gives error bounds for the oracle estimator.

\begin{lemma}
\label{LEM:oracle}
Under Assumptions (A1) -- (A6) in the Supplementary Materials, the oracle estimator satisfies the following error bounds, 
\begin{align*}
& \| \widehat{\bs{\xi}}_{t}^o - [\bs{\xi}_{t}]_{\mathcal{N}_t}\|_{2} = O_P\left( |\mathcal{N}_t|^{1/2}t_0^{-1/2}\right),\\
&\sum_{k=1}^p\|\widehat{\alpha}^o_{kt} - \alpha_{kt}\|_{L_2}  + \|\widehat{\beta}^o_t - \beta_t\|_{L_2} = O_P\left\{H^{\varrho}+|\triangle|^{d+1} + n^{-1/2} (|\triangle|^{-1}+H^{-1/2}) + \lambda_0 n^{-1} t_0^{-1}|\triangle|^{-4}\right\},
\end{align*}
where $|\triangle|:=\max \{|\tau|,\tau \in \triangle \}$ is the size of $\triangle$, $H$ is the distance between two consecutive knots, $[\bs{\xi}_{t}]_{\mathcal{N}_t}$ is the vector obtained by taking the elements of $\bs{\xi}_t$ at the non-zero index set, $|\mathcal{N}_t|$ is the cardinality of set $\mathcal{N}_t$, and $\|\cdot\|_2$ is the Euclidean norm, and $\|f\|_{L_2}=\int_{\bs{u}\in \Omega} f(\bs{u})^2d\bs{u}$ is the $L_2$-norm of the function $f$.
\end{lemma}

Lemma \ref{LEM:oracle} states that the estimation errors of the oracle estimator depend on the triangle size $|\triangle|$, distance between two consecutive knots of univariate spline $H$, data size $n$, and tuning parameter $\lambda_0$.
Setting $|\triangle|^{-1} = o(\sqrt{n})$, $H^{-1} = o(n)$, and tuning parameter $\lambda_0 = o(nt_0|\triangle|^{4})$ ensures that the oracle estimator is consistent as $n \to \infty$.
Next, we show in Theorem \ref{THE:consistency} that under mild conditions, the \texttt{RST-GAM} estimator is also consistent. 
The key idea is that the oracle estimator satisfies the Karush-Kuhn-Tucker (KKT) conditions of  problem \eqref{eq:Model_Estimation_4} almost surely. 
Consequently, our proposed estimator coincides with the oracle one and is therefore consistent.

\begin{theorem}
\label{THE:consistency}
Under Assumptions (A1) -- (A7) in the Supplementary Materials, the \texttt{RST-GAM}  estimator satisfies the following error bounds,
\begin{align*}
& \| \widehat{\bs{\xi}}_t - \bs{\xi}_t \|_2 = O_P\left(|\mathcal{N}_t|^{1/2}t_0^{-1/2}+ \lambda_1 \|[\V{w}_t]_{\mathcal{N}_t}\|_2\right),\\ 
&\|\widehat{\alpha}_{kt} - \alpha_{kt}\|_{L_2}= O_P\left\{H^{\varrho}+|\triangle|^{d+1} + n^{-1/2} (|\triangle|^{-1}+H^{-1/2}) + \lambda_0 n^{-1} t_0^{-1}|\triangle|^{-4} + \lambda_1 \|[\V{w}_t]_{\mathcal{N}_t}\|_2\right\}, \\
&\|\widehat{\beta}_t - \beta_t\|_{L_2}=O_P\left\{H^{\varrho}+|\triangle|^{d+1} + n^{-1/2} (|\triangle|^{-1}+H^{-1/2}) + \lambda_0 n^{-1} t_0^{-1}|\triangle|^{-4} + \lambda_1 \|[\V{w}_t]_{\mathcal{N}_t}\|_2\right\}, 
\end{align*}
where $|\mathcal{N}_t|$ is the cardinality of the set $\mathcal{N}_t$, $[\V{w}_t]_{\mathcal{N}_t}$ is a subvector of $\V{w}_t$, and $\|[\V{w}_t]_{\mathcal{N}_t}\|_2 = (\sum_{i \in \mathcal{N}_t} w_{it}^2)^{1/2}$. Furthermore, the slack variable estimator $\widehat{\bs{\xi}}_t$ is selection consistent, i.e., as $n \to \infty$,
\begin{eqnarray*}
\mathbb{P}\left( \widehat{\mathcal{N}}_t = {\mathcal{N}}_t \right) & \to &  1,
\end{eqnarray*}
where $ \widehat{\mathcal{N}}_t = \left\{i : \widehat{\xi}_{it} \neq 0\right\}$ is the estimated non-zero index set of $\bs{\xi}_t$. 
\end{theorem}

Theorem \ref{THE:consistency} states that, like the error bounds on the oracle estimator, the error bounds on the \texttt{RST-GAM}  estimator depend on the triangle size $|\triangle|$ and the distance between interior knots $H$, data size $n$, and tuning parameter $\lambda_0$. Additionally, they also depend on the tuning parameter $\lambda_1$ and the term $\|[\V{w}_t]_{\mathcal{N}_t}\|_2$, which implies that when the weights $w_{it}$ for the slack variables corresponding to outliers are small, the overall Lasso penalty will induce less bias. By setting $H$, $\lvert \triangle \rvert$, and $\lambda_0$, as we did to ensure consistency of the oracle estimator, and further setting $\lambda_1 = o(\|[\V{w}_t]_{\mathcal{N}_t}\|_2^{-1})$, we ensure  estimation consistency of the \texttt{RST-GAM} estimator. Proofs of Lemma \ref{LEM:oracle} and Theorem \ref{THE:consistency} are provided in Section 3 of the Supplementary Materials.

\section{Estimation Algorithm}
\label{sec:algorithm}
We estimate the \texttt{RST-GAM} parameters, which minimize the convex objective function in 
\eqref{eq:Model_Estimation_1}, via proximal gradient descent (PGD). The PGD algorithm is a natural choice for minimizing the sum of two functions $f(\V{z}) + g(\V{z})$ when $f(\V{z})$ is a differentiable function and $g(\V{z})$ is a nonsmooth function whose proximal map can be evaluated analytically or admits efficient computation \citep{ComWaj2005, combettes2011proximal,  parikh2014proximal}.  The proximal map of $g(\V{z})$ is given by
\begin{eqnarray*}
\prox_{\eta g}(\V{z}) & = & \underset{\V{\zeta}}{\arg\min}\; \left\{g(\V{\zeta}) + \frac{1}{2\eta}\lVert \V{\zeta} - \V{z}\rVert^2\right\}
\end{eqnarray*}
where $\eta$ is a positive parameter. It exists and is unique whenever $g(\V{z})$ is convex and lower semicontinuous. 

Given the $k$th iterate $\Vn{z}{k}$, the PGD algorithm computes the next iterate by taking a gradient step followed by applying a proximal map.
\begin{eqnarray}
    \Vn{z}{k+1} &=& \prox_{\eta_k g} \left( \Vn{z}{k} - \eta_k \nabla f\left(\Vn{z}{k}\right) \right).
\label{eq:proximal_update}
\end{eqnarray}
The positive step-size parameter $\eta_k$ may vary with the iteration. Larger $\eta_k$ can lead to more progress per-iteration and faster convergence. But if $\eta_k$ is too large, the iterate sequence will diverge. The procedure for setting $\eta_k$ is critical to ensuring the convergence of PGD. 

We use the adaptive procedure proposed by \citet{malitsky2024adaptive} for selecting $\eta_k$, since it comes with convergence guarantees when we apply PGD to 
the \texttt{RST-GAM} estimation problem in \eqref{eq:Model_Estimation_3}. Algorithm~\ref{alg:[gd]} provides pseudocode for general PGD using their adaptive procedure for setting $\eta_k$. Section 3 in the supplements provides details on the adaptive procedure, including initialization of $\eta_0$ and a hyperparameter $\upsilon_0$, as well as an explanation for why we use this adaptive procedure for choosing $\eta_k$, rather than a simpler backtracking line search, to fit the \texttt{RST-GAM} model. 

\begin{algorithm} [H]
Initialize $\Vn{z}{0} \gets \left\{\V{\gamma}_t^{*(0)}, \Vn{\theta}{0}_t, \Vn{\xi}{0}_t\right\}, \upsilon_0, \eta_0$
    \caption{Proximal Gradient Descent with  Adaptive Step-Size} 
    \label{alg:[gd]}
    \begin{algorithmic}
\State for $k=1, 2, \ldots$
            \Repeat 
                \State $L_k \gets \frac{\left\lVert \nabla f\left(\Vn{z}{k}\right) - \nabla f\left(\Vn{z}{k-1}\right)\right\rVert_2}{\left\lVert \Vn{z}{k} - \Vn{z}{k-1}\right\rVert_2}$
                \Comment{Estimate local Lipschitz constant}
                \State $\eta_k \gets \min \left\{ \sqrt{2/3 + \upsilon_{k-1}} \eta_{k-1}, \frac{\eta_{k-1}}{\sqrt{\left[2 \eta_{k-1}^2 L_k^2 - 1\right]_+}}\right\}$         \Comment{Update step-size}
                \State $\upsilon_k \gets \frac{\eta_k}{\eta_{k-1}}$
                \State $\Vn{z}{k+1} \gets \prox_{\eta_k g}\left(\Vn{z}{k} - \eta_k \nabla f\left(\Vn{z}{k}\right)\right)$ \;
		\Until{convergence}
    \end{algorithmic} 
\end{algorithm}

To apply Algorithm~\ref{alg:[gd]} to estimate the \texttt{RST-GAM} parameters, observe that the objective function in \eqref{eq:Model_Estimation_3} is the sum of the two functions
\begin{align}
f(\V{\gamma}_t^*, \V{\theta}_t, \V{\xi}_t) 
  &\amp = \amp -\sum_{i=1}^{n}\sum_{s=t-t_0}^{t}  
      L\left[\exp\left\{\M{B}(\V{u}_i){\Tra}\M{Q}_2\V{\gamma}_t^*
         + \sum_{k=1}^{p}\M{A}_k(X_{isk}){\Tra}\V{\theta}_{kt}
         + \xi_{it}\right\}, 
         Y_{is}\right]
      \nonumber\\
  &\quad
     + \frac{\lambda_{0}}{2}\,
       (\V{\gamma}_t^*){\Tra}\,
        \M{Q}_2{\Tra}\,\M{P}\,\M{Q}_2\,
       \V{\gamma}_t^*
       \label{eq:smooth_part_f} \\[1em]
g(\V{\gamma}_t^*, \V{\theta}_t, \V{\xi}_t) 
  & \amp = \amp \tilde{g}\left(\V{\xi}_t\right) \amp = \amp 
  \lambda_1 \sum_{i=1}^n w_{it} \,\lvert \xi_{it} \rvert + \iota_{\Real_+}\left(\V{\xi}_t\right)
     \label{eq:nonsmooth_part_g},
\end{align}
where $\iota_{\Real_+}(\V{\xi})$ is the indicator function of the nonnegative orthant, i.e., $\iota_{\Real_+}(\V{\xi})$ is zero when $\V{\xi} \in \Real_+$ and is infinity otherwise. The proximal map of $g$ in \eqref{eq:nonsmooth_part_g} is given by
\begin{eqnarray}
\prox_{\gamma \tilde{g}}\left(
\V{\gamma}, \V{\theta}, \V{\xi} \right) & = & \left\{
\V{\gamma}, \V{\theta}, \prox_{\gamma \tilde{g}}(\V{\xi})
\right\},
\label{eq:soft_thresholding_new}
\end{eqnarray}
where the $i$th element of $\prox_{\gamma \tilde{g}}(\V{\xi})$ is the composition of the softthresholding operator and the  projection onto the nonnegative orthant.
\begin{eqnarray*}
\left[\prox_{\gamma \tilde{g}}\left(
\V{\xi} \right)\right]_i & = & 
[\xi_{i}- \lambda_1 w_{i}]_+ \quad\quad \text{for $i \in [n]$}
\end{eqnarray*}

Applying Algorithm~\ref{alg:[gd]} to \texttt{RST-GAM} model estimation \eqref{eq:Model_Estimation_4} comes with the following convergence guarantees.

\begin{proposition}
\label{prop:PGD_convergence}
Let $\Vn{z}{k} = \left\{\V{\gamma}_t^{*(k)}, \Vn{\theta}{k}_t, \Vn{\xi}{k}_t\right\}$ denote the iterates generated by Algorithm~\ref{alg:[gd]} with 
$f(\V{z})$ and $g(\V{z})$ defined in \eqref{eq:smooth_part_f} and \eqref{eq:nonsmooth_part_g} respectively.
Let $F(\V{z}) = f(\V{z}) + g(\V{z})$ and $F^*$ be the value of the objective function in \eqref{eq:Model_Estimation_4} at its global minimizer. Then the iterates $\Vn{z}{k}$ converge to a solution of \eqref{eq:Model_Estimation_4} and 
    \begin{eqnarray*}
    \min \left\{F\left(\Vn{z}{1}\right), F\left(\Vn{z}{2}\right), \cdots, 
        F\left(\Vn{z}{k}\right)  \right\} & \le & F^* + \frac{C}{k}
    \end{eqnarray*}
for some positive $C$.
\end{proposition}
\begin{proof}
See Section 3 of the Supplementary Materials.
\end{proof}

Proposition~\ref{prop:PGD_convergence} guarantees the convergence of Algorithm~\ref{alg:[gd]} as well as 
bounds the rate at which the objective function converges to its optimal value. We can use the optimality conditions of \eqref{eq:Model_Estimation_4} as a practical check for convergence.
\begin{proposition}
\label{prop:KKT}
Let
\begin{eqnarray*}
h(\V{\gamma}_t^*, \V{\theta}_{t}, \V{\xi}_t) & = & \M{B}(\V{u}_i){\Tra}\M{Q}_2\V{\gamma}_t^* + \sum_{k=1}^{p}\M{A}_k(X_{isk}){\Tra}\V{\theta}_{kt} + \xi_{it}.
\end{eqnarray*}
A point $(\V{\gamma}_t^*, \V{\theta}_{t}, \V{\xi}_t)$ is the global minimizer to the optimization problem \eqref{eq:Model_Estimation_4} if and only if it satisfies the following conditions.
\begin{enumerate}[(a)]
\item $\xi_{it} \;\ge\; 0,$ \quad for $i \in [n]$
\item $\sum_{i=1}^{n}\sum_{s=t-t_0}^{t} \frac{\exp\{h(\V{\gamma}_t^*, \V{\theta}_{t}, \V{\xi}_t)\} - Y_{is}}{\sigma^2} \M{Q}_2{\Tra}\M{B}(\V{u}_i) + \lambda_0 \M{Q}_2{\Tra} \M{P} \M{Q}_2 \V{\gamma}_t^* \amp = \amp \V{0}$
\item $\sum_{i=1}^{n}\sum_{s=t-t_0}^{t} \frac{\exp\{h(\V{\gamma}_t^*, \V{\theta}_{t}, \V{\xi}_t)\} - Y_{is}}{\sigma^2} \M{A}_k(X_{isk}) \amp = \amp \V{0}, \quad \text{for $k \in [p]$}$
\item $\sum_{s=t-t_0}^{t} \frac{\exp\{h(\V{\gamma}_t^*, \V{\theta}_{t}, \V{\xi}_t)\} - Y_{is}}{\sigma^2} + \nu_{it} \amp = \amp 0\,$,  
$\quad \nu_{it} \;\in\;
\begin{cases}
\{\lambda_1\,w_{it}\}, & \text{if }\xi_{it}>0,\\[4pt]
[\,0,\;\lambda_1\,w_{it}\,], & \text{if }\xi_{it}=0.
\end{cases} \quad \text{for $i \in [n]$}$
\end{enumerate}
\end{proposition}
\begin{proof}
See Section 3 of the Supplementary Materials.
\end{proof}

Finally, to expedite computations, we can accelerate Algorithm~\ref{alg:[gd]} using momentum-based strategies.  
We use this accelerated PGD algorithm for all our numerical studies and rely on the optimality conditions in Proposition~\ref{prop:KKT} to check for convergence. We provide details on the accelerated PGD algorithm in Section 3 of the Supplementary Materials. 



\section{Model Selection}
\label{sec:model_selection}
Selecting the tuning parameters $\lambda_0$ and $\lambda_1$ in problem (\ref{eq:Model_Estimation_4}) presents a challenge. A brute-force grid search over both parameters is computationally impractical. Consequently, we adopt a two-stage selection procedure \citep{Zhang2024}. Suppose we have two sets of candidate values for $\lambda_0$ and $\lambda_1$. In the first stage, we set  $\lambda_0 = 0$ and select the $\lambda_1$ candidate that minimizes the Bayesian information criterion (BIC)
\begin{eqnarray}
    \textbf{BIC}_{\lambda_1}\left(\widehat{\V{\gamma}}_t^*, \widehat{\V{\theta}}_t, \widehat{\V{\xi}}_t\right) & = & -2\,\ell \left(\mu\left(\widehat{\V{\gamma}}_t^*, \widehat{\V{\theta}}_t, \widehat{\V{\xi}}_t\right), \M{Y} \right) + \log(N) \df(\widehat{\V{\xi}}_t)
\label{BIC lasso selection}
\end{eqnarray}
where $\ell(\boldsymbol{\mu}, \M{Y})$ is the Poisson log-likelihood, $N = n(t_0 + 1)$ is the total number of data points, and $\df(\widehat{\V{\xi}}_t)$ is the number of non-zero elements in $\widehat{\V{\xi}}_t$. Let $\lambda_1^*$ denote the Lasso penalty that minimizes (\ref{BIC lasso selection}). In the second stage, we set $\lambda_1 = \lambda_1^*$  and select the $\lambda_0$ candidate that minimizes the extended Bayesian information criterion (EBIC) proposed by \citet{chen2008extended}.
\begin{align}
    \textbf{EBIC}_{\lambda_0}\left(\widehat{\V{\gamma}}_t^*, \widehat{\V{\theta}}_t, \widehat{\V{\xi}}_t\right) = -2\,\ell \left( \mu\left(\widehat{\V{\gamma}}_t^*, \widehat{\V{\theta}}_t, \widehat{\V{\xi}}_t\right), \M{Y} \right) &+ \log(N) \,\df(\widehat{\V{\gamma}}_t^*, \widehat{\V{\theta}}_t, \widehat{\V{\xi}}_t) \nonumber \\
    &+ 2 \rho \log  {
    \binom{P}{\df(\widehat{\V{\gamma}}_t^*, \widehat{\V{\theta}}_t, \widehat{\V{\xi}}_t)}
    },
\label{EBIC roughness selection}
\end{align}
where $\rho \in [0,1]$ is a hyperparameter and $P$ is the number of parameters in the model. We detail how to compute $\df(\widehat{\V{\gamma}}_t^*, \widehat{\V{\theta}}_t, \widehat{\V{\xi}}_t)$ in Section 4 of supplements.

\section{Numerical Studies}
\label{sec:simulation}
In this section, we present two numerical studies that evaluate the finite-sample performance of our \texttt{RST-GAM} method across various scenarios. The first study highlights its ability to provide robust estimates while simultaneously detecting outliers under varying outlier intensities and quantities over a simple spatial domain. The second study assesses its robustness in a more complex scenario using synthetic pandemic data.

As a baseline comparison, we fit a spatiotemporal generalized additive model (\texttt{NST-GAM}) that omits the slack variables to quantify the gains achieved by explicitly modeling outliers with $\xi_{it}$. Specifically, the \texttt{NST-GAM} estimates parameters by solving the optimization problem
\begin{equation}
        \underset{\V{\gamma}_t^*, \V{\theta}_{kt}}{\text{minimize}} -\sum_{i=1}^{n}\sum_{s=t-t_0}^{t} L\left[\exp\left\{\M{B} (\V{u}_i){\Tra}\M{Q}_2\V{\gamma}^*_t+ \sum_{k=1}^{p}\M{A}_k(X_{isk}){\Tra}\V{\theta}_{kt} \right\}, Y_{is}\right] + \frac{\lambda}{2} (\V{\gamma}^*_t){\Tra}\M{Q}_2^{\Tra}\M{P}\M{Q}_2\V{\gamma}^*_t.
\label{eq:Model_without_slack}
\end{equation}
\texttt{NST-GAM} extends the generalized geoadditive models (GGAMs) of \citep{Yu2020} from purely spatial to spatiotemporal data. We compute estimates for $\hat{\V{\gamma}}^*_t$ and $\hat{\V{\theta}}_{kt}$ using Algorithm 2 by excluding all $\xi_{it}$ terms and setting $g(\V{\gamma}_t^*, \V{\theta}) = 0$.

\subsection{Varying Outlier Cases over a Horseshoe Domain Domain}
\label{subsec:exp_1}
\begin{figure}[H]
\captionsetup{font=small}
    \centering
    \subfloat[\centering Univariate Components $\alpha_{kt}$]{{\includegraphics[height = 1.6in, width=6.36in]{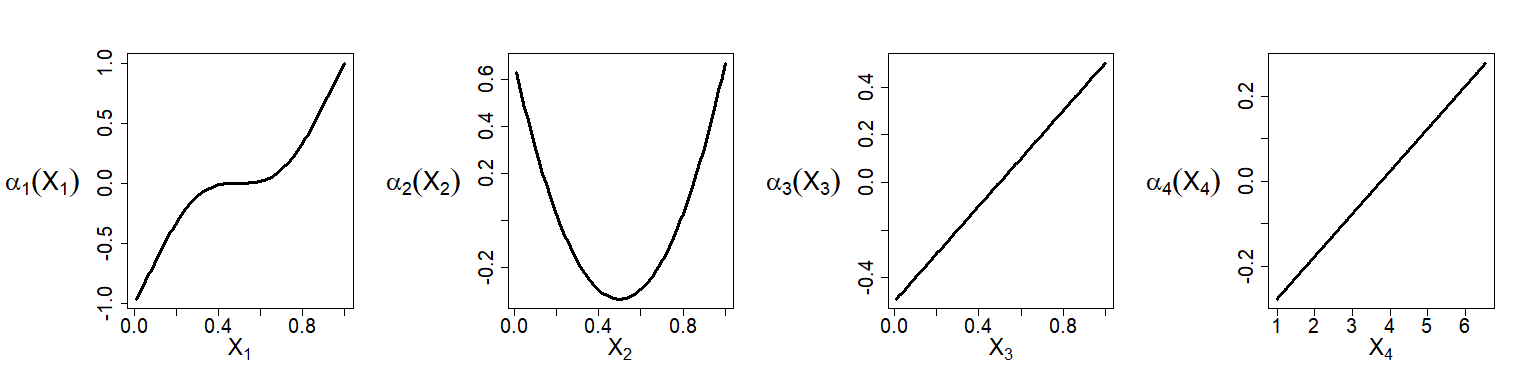} }}%
    \qquad
    \vspace{3mm}
    \subfloat[\centering Bivariate Component $\beta_t$]{{\includegraphics[height = 1.8in, width=2.8in]{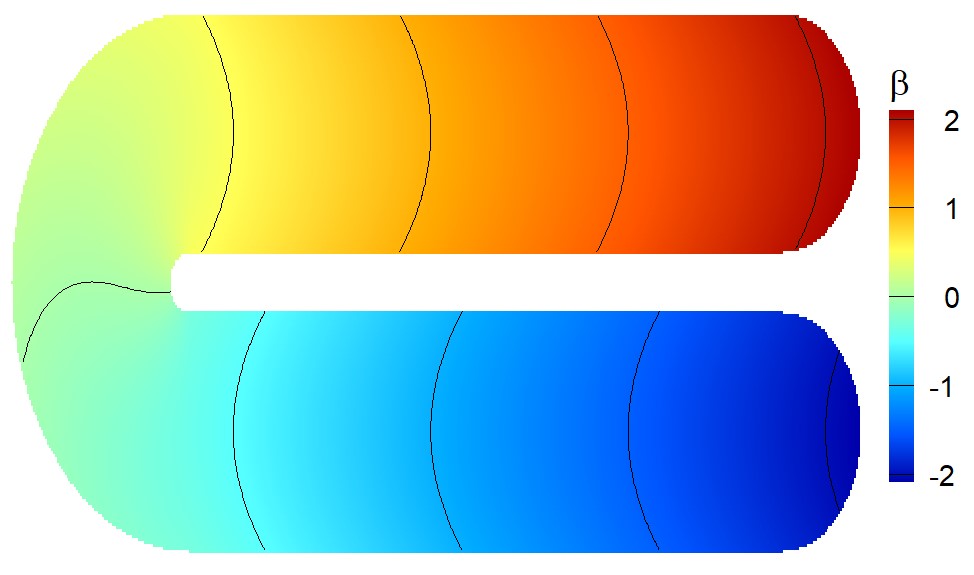} }}%
    \qquad
    \subfloat[\centering Triangulation Plot]{{\includegraphics[height = 1.8in, width=2.6in]{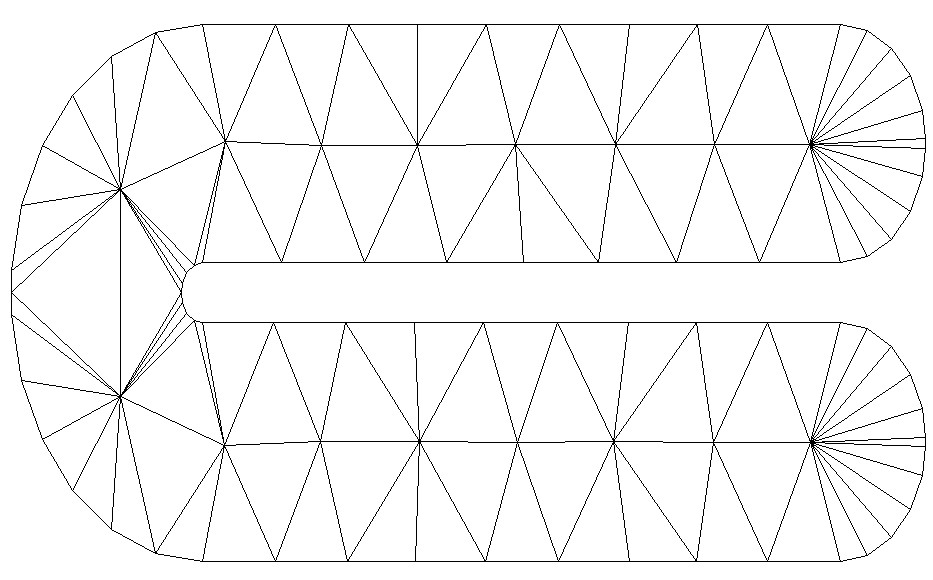} }}%
    \caption{Simulated true functional forms and triangulation in numerical study 1}
    \label{fig:Example_1_univariate_and_bivariate_Comp}%
\end{figure}

We first evaluate \texttt{RST-GAM}'s resistance to outliers as their magnitude and number vary. This numerical study uses the same  data generation procedure in Example 1 of \cite{Yu2020} with one modification: the univariate component has an extra time-dependent feature. The number of counts $Y_{it}$ at location $i \in [2000]$ and time $t \in [5]$ follows a Poisson distribution, i.e., $Y_{it} \sim \text{Poisson}(\mu_{it})$, where  $\mu_{it} = \exp\left(\beta_{t}(\V{u}_i) + \sum_{k=1}^{4}\alpha_{kt}(X_{itk})\right)$. The location coordinates $\V{u}_i$ are uniformly sampled over a horseshoe-shaped domain \citep{horseshoe2013}, and the covariates $\{X_{itk} \}_{k=1}^3 = \{X_{ik} \}_{k=1}^3$ are time-independent iid samples from $\text{Unif}(0,1)$. The fourth covariate $X_{it4}$, however, is time-dependent and generated as $X_{it4} = \log\left(\sum_{s = 1}^{t-1} Y_{is}\right)$ for $t = 2, \dots, 5$ with an initial value of 1. We construct $X_{it4}$ this way to mimic the effect of accumulated infection counts on new infections at time $t$. Figures~\ref{fig:Example_1_univariate_and_bivariate_Comp} (a) and (b) display the functional forms for the univariate components $\{\alpha_{kt} \}_{k=1}^4$ and the bivariate component $\beta_t$, respectively.

We introduce outliers by adding a constant positive shift to the underlying mean parameter $\mu_{it}$ at randomly selected locations, i.e., $\xi_{it} > 0$, over the time period $t$. We evaluate the estimation robustness and outlier detection accuracy of the proposed method with two simulation designs:

\begin{itemize}
    \item \textbf{Varying Outlier Strength:} We fix the number of outliers as $50$ and vary the magnitude of the positive shift to $\mu_{it}$. Shift magnitudes come from the set $\{1, 5, 15, 30, 50, 100\}$.
    \item \textbf{Varying Outlier Quantity:} We fix the shift magnitude at $30$ and vary the number of outlier, i.e., number of location points $i$. The numbers come from the set $\{1, 10, 50, 100, 150, 200\}$.
\end{itemize}

We generate 100 Monte Carlo (MC) replicates with the same sample size of $n = 2000$ and $t \in [5]$ for both designs. To construct the adaptive Lasso weights, we use $Q = 2$ data-thinning folds. To estimate the univariate functions in (\ref{eq:univariate_basis}), we use cubic splines with 4 interior knots. To estimate the bivariate component in (\ref{eq:bivariate_basis}), we use a triangulation, consisting of 109 triangles and 95 vertices, using the R package ``Triangulation" \citep{Triangulation}. Figure~\ref{fig:Example_1_univariate_and_bivariate_Comp} (c) plots the resulting triangulation. In general, the quality of bivariate estimation saturates with sufficiently fine triangulation, but the point of diminishing returns depends on the structure of the spatial domain \citep{lai2007,Yu2020}. We advise users to balance triangulation resolution with spatial domain complexity to maximize returns on computational effort.

\begin{figure}[H]
\captionsetup{font=small}
    \centering
    \subfloat[\centering Bivariate MISE versus outlier strength]{{\includegraphics[height = 1.8in, width=2.95in]{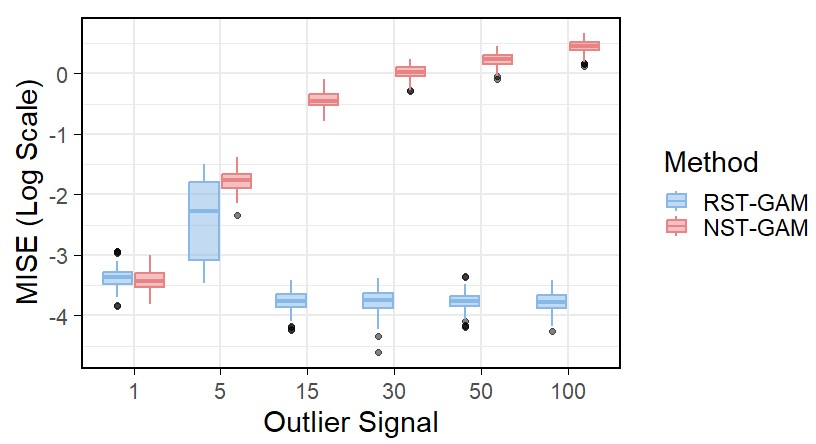} }}%
    \qquad
    \subfloat[\centering Bivariate MISE versus outlier quantity]{{\includegraphics[height = 1.8in, width=2.95in]{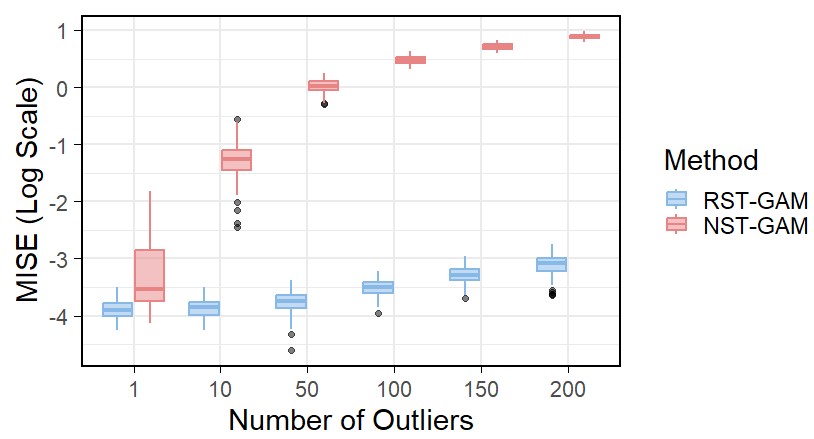} }}%
    \caption{Boxplots of bivariate estimation MISE of \texttt{RST-GAM} under varying outlier strength and quantity}%
    \label{fig:MC_Bivariate}%
\end{figure}

We quantify estimation accuracy with the mean integrated squared error (MISE), i.e.,
\begin{eqnarray*}
\text{MISE}\left(\hat{\beta}_t\right) & = & \frac{1}{n}\sum_{i=1}^n \left(\hat{\beta}_t(\V{u}_i) - \beta_t(\V{u}_i)\right)^2 \\
\text{MISE}\left(\hat{\alpha}_{kt}\right) & = &  \frac{1}{n}\sum_{i=1}^n \left(\hat{\alpha}_{kt}(X_{itk}) - \alpha_{kt}(X_{itk})\right)^2.
\end{eqnarray*}
Table~\ref{table:MISE_functional_component_estimators} reports the average MISE over the 100 replicates. \texttt{RST-GAM} achieves comparable or lower MISE values than the standard
 approach (\ref{eq:Model_without_slack}). Moreover, \texttt{RST-GAM}'s estimation accuracy is robust over a wide range of outlier shift magnitudes and outlier quantities. Figure~\ref{fig:MC_Bivariate} (a) and (b) displays boxplots of the MISE on a logarithmic scale for the bivariate component under varying outlier strength and outlier quantity respectively. The boxplots of MISE for each estimated univariate component are provided in the Supplemental Materials.

\vspace{2mm}

We evaluate outlier detection performance using the false positive rate (FPR) and false negative rate (FNR). Recall that a sample $Y_{it}$ is flagged as an outlier if its slack estimate $\hat{\xi}_{it}$ is positive.  The last two columns in Table~\ref{table:MISE_functional_component_estimators} report the average FPR and FNR. Ideally, the FNR should be zero to avoid missing any potential outliers, while the FPR should be low to minimize the incorrect identification of non-outliers. Table~\ref{table:MISE_functional_component_estimators} tells us that  \texttt{RST-GAM} consistently achieves low FPR across all scenarios and zero FNR except when outlier strength is low, in which case the outliers have marginal impact on parameter estimation. These findings demonstrate that \texttt{RST-GAM}  can identify influential outliers accurately.

\vspace{2mm}
\begin{table}[H]
\captionsetup{font=small}
\centering
\setlength{\tabcolsep}{10pt}
\renewcommand{\arraystretch}{1.8}
\caption{MISE of the bivariate and univariate function estimates and FPR and FNR for outlier detection over 100 replicates}
\large
\resizebox{\textwidth}{!}{
\begin{tabular}{llp{1.1cm}p{1.1cm}p{1.1cm}p{1.1cm}p{1.1cm}p{1.1cm}p{1.1cm}p{1.1cm}p{1.1cm}p{1.1cm}cc}
\hline
\multirow{3}{*}{\textbf{Simulation Case}} & \multirow{3}{*}{\textbf{Size}} & \multicolumn{10}{c}{\textbf{MISE}} & \multirow{3}{*}{\textbf{FPR}} & \multirow{3}{*}{\textbf{FNR}} \\ \cline{3-12}
                         &               & \multicolumn{5}{c}{\textbf{\texttt{RST-GAM}}} & \multicolumn{5}{c}{\textbf{\texttt{NST-GAM}}} \\ \cline{3-7} \cline{8-12}
                         &               & \(\boldsymbol{\beta_t}\) & \(\boldsymbol{\alpha_{1t}}\) & \(\boldsymbol{\alpha_{2t}}\) & \(\boldsymbol{\alpha_{3t}}\) & \(\boldsymbol{\alpha_{4t}}\) & \(\boldsymbol{\beta_t}\) & \(\boldsymbol{\alpha_{1t}}\) & \(\boldsymbol{\alpha_{2t}}\) & \(\boldsymbol{\alpha_{3t}}\) & \(\boldsymbol{\alpha_{4t}}\) \\ \hline

\multirow{6}{*}{Outlier Strength}
& 1   & \smallnum{3.5e-2} & \smallnum{5.4e-4} & \smallnum{2.8e-4} & \smallnum{3.5e-4} & \smallnum{6.6e-4} & \smallnum{3.4e-2} & \smallnum{5.1e-4} & \smallnum{2.7e-4} & \smallnum{3.4e-4} & \smallnum{6.0e-4} & \smallnum{0} & \smallnum{1} \\ \cline{2-14}
& 5   & \smallnum{1.1e-1} & \smallnum{2.7e-3} & \smallnum{1.2e-3} & \smallnum{1.3e-3} & \smallnum{6.5e-3} & \smallnum{1.7e-1} & \smallnum{4.6e-3} & \smallnum{1.9e-3} & \smallnum{2.0e-3} & \smallnum{1.2e-2} & \smallnum{4.2e-4} & \smallnum{6.7e-1} \\ \cline{2-14}
& 15  & \smallnum{2.4e-2} & \smallnum{4.5e-4} & \smallnum{2.8e-4} & \smallnum{2.9e-4} & \smallnum{5.2e-4} & \smallnum{6.5e-1} & \smallnum{4.8e-2} & \smallnum{2.3e-2} & \smallnum{2.2e-2} & \smallnum{3.2e-1} & \smallnum{1.5e-4} & \smallnum{0} \\ \cline{2-14}
& 30  & \smallnum{2.4e-2} & \smallnum{4.5e-4} & \smallnum{2.7e-4} & \smallnum{3.0e-4} & \smallnum{3.7e-4} & \smallnum{1.0e+0} & \smallnum{1.0e-1} & \smallnum{4.9e-2} & \smallnum{4.4e-2} & \smallnum{9.2e-1} & \smallnum{1.2e-4} & \smallnum{0} \\ \cline{2-14}
& 50  & \smallnum{2.4e-2} & \smallnum{4.4e-4} & \smallnum{2.7e-4} & \smallnum{3.0e-4} & \smallnum{2.5e-4} & \smallnum{1.2e+0} & \smallnum{1.3e-1} & \smallnum{6.5e-2} & \smallnum{5.8e-2} & \smallnum{1.4e+0} & \smallnum{1.9e-4} & \smallnum{0} \\ \cline{2-14}
& 100 & \smallnum{2.3e-2} & \smallnum{4.4e-4} & \smallnum{2.9e-4} & \smallnum{3.4e-4} & \smallnum{1.9e-4} & \smallnum{1.6e+0} & \smallnum{1.5e-1} & \smallnum{7.9e-2} & \smallnum{6.9e-2} & \smallnum{2.3e+0} & \smallnum{1.3e-4} & \smallnum{0} \\ \hline

\multirow{6}{*}{Outlier Quantity}
& 1   & \smallnum{2.1e-2} & \smallnum{4.3e-4} & \smallnum{2.3e-4} & \smallnum{3.3e-4} & \smallnum{4.3e-4} & \smallnum{5.0e-2} & \smallnum{1.5e-3} & \smallnum{6.9e-4} & \smallnum{1.0e-3} & \smallnum{8.5e-3} & \smallnum{1.0e-4} & \smallnum{0} \\ \cline{2-14}
& 10  & \smallnum{2.1e-2} & \smallnum{4.4e-4} & \smallnum{2.4e-4} & \smallnum{3.3e-4} & \smallnum{4.7e-4} & \smallnum{3.1e-1} & \smallnum{2.1e-2} & \smallnum{1.1e-2} & \smallnum{1.1e-2} & \smallnum{3.1e-1} & \smallnum{1.0e-4} & \smallnum{0} \\ \cline{2-14}
& 50  & \smallnum{2.4e-2} & \smallnum{4.5e-4} & \smallnum{2.7e-4} & \smallnum{3.0e-4} & \smallnum{3.7e-4} & \smallnum{1.0e+0} & \smallnum{9.9e-2} & \smallnum{4.9e-2} & \smallnum{4.4e-2} & \smallnum{9.2e-1} & \smallnum{1.2e-4} & \smallnum{0} \\ \cline{2-14}
& 100 & \smallnum{3.0e-2} & \smallnum{4.9e-4} & \smallnum{2.7e-4} & \smallnum{3.8e-4} & \smallnum{5.4e-4} & \smallnum{1.6e+0} & \smallnum{1.4e-1} & \smallnum{6.6e-2} & \smallnum{6.1e-2} & \smallnum{1.1e+0} & \smallnum{4.1e-4} & \smallnum{0} \\ \cline{2-14}
& 150 & \smallnum{3.7e-2} & \smallnum{5.9e-4} & \smallnum{3.1e-4} & \smallnum{3.8e-4} & \smallnum{8.3e-4} & \smallnum{2.0e+0} & \smallnum{1.6e-1} & \smallnum{7.4e-2} & \smallnum{6.8e-2} & \smallnum{1.2e+0} & \smallnum{4.0e-4} & \smallnum{0} \\ \cline{2-14}
& 200 & \smallnum{4.5e-2} & \smallnum{6.6e-4} & \smallnum{3.5e-3} & \smallnum{4.2e-4} & \smallnum{1.2e-3} & \smallnum{2.4e+0} & \smallnum{1.7e-1} & \smallnum{7.8e-2} & \smallnum{7.1e-2} & \smallnum{1.3e+0} & \smallnum{6.4e-4} & \smallnum{0} \\ \hline

\end{tabular}
}
\label{table:MISE_functional_component_estimators}
\end{table}

\vspace{-5pt}
\subsection{Synthetic Infections Counts Based on COVID-19 Covariates}
\label{subsec:exp_2}
\begin{figure}[H]
\captionsetup{font=small}
    \centering
    \subfloat[\centering Simulated $\beta_\text{true}$]{{\includegraphics[height = 1.6in, width=3in]{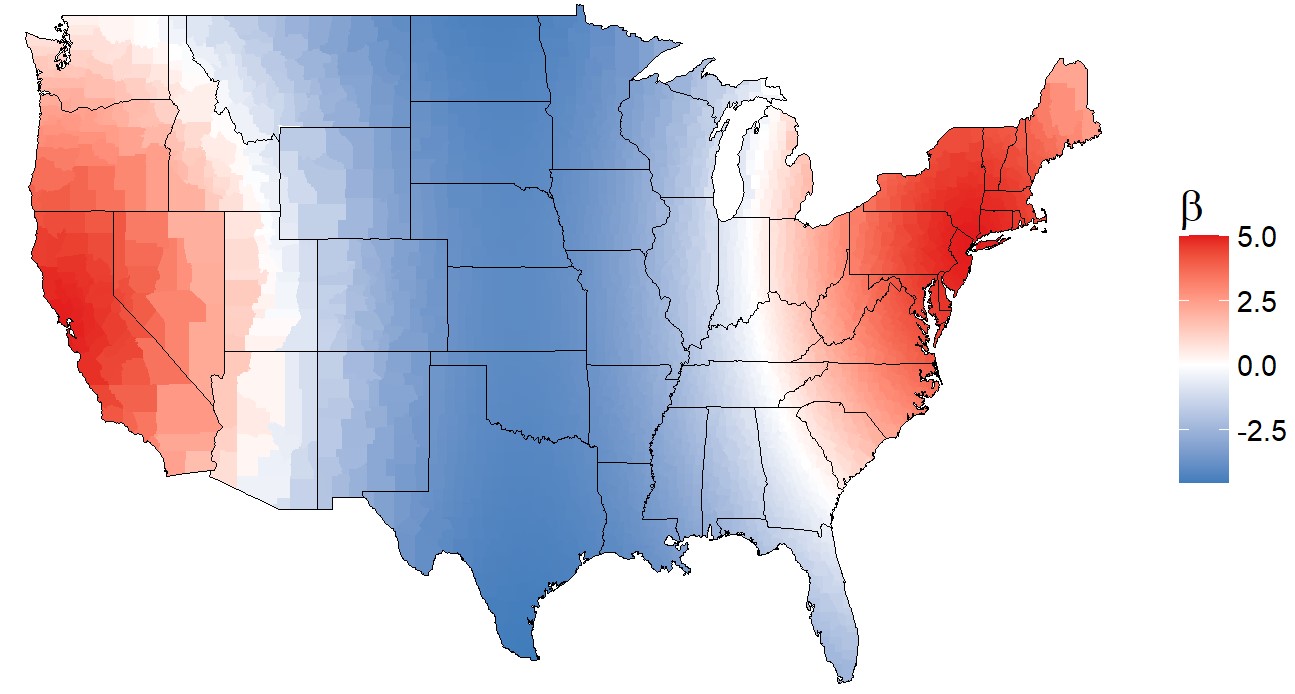} }}%
    \qquad
    \subfloat[\centering Simulated Outlier Location]{{\includegraphics[height = 1.6in, width=2.8in]{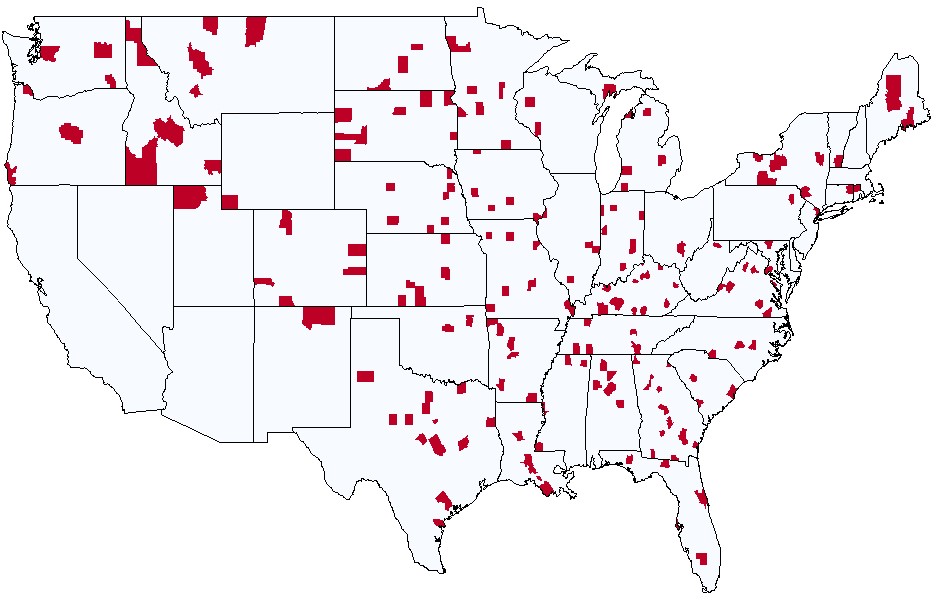} }}%
\vspace{0.5cm} 
    \subfloat[\centering Estimated $\widehat{\beta}_\text{\texttt{RST-GAM}}$ (MISE: $0.74$)]{{\includegraphics[height = 1.6in, width=3in]{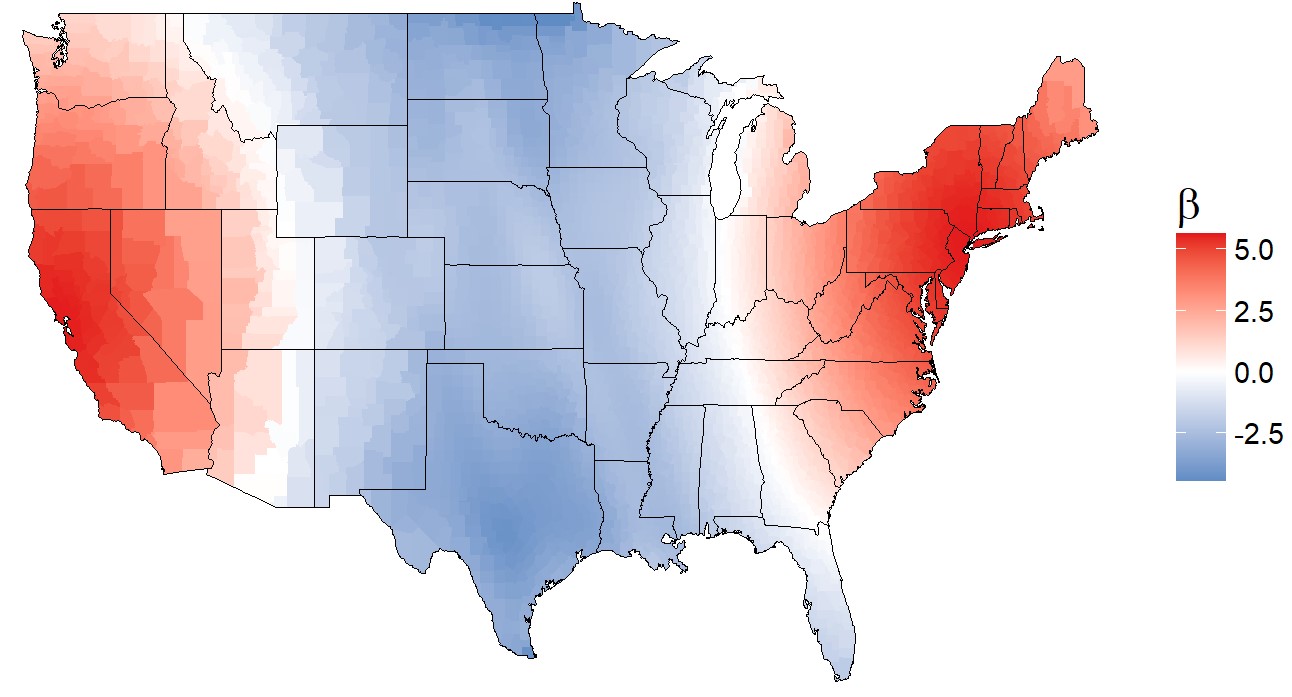} }}%
    \qquad
    \subfloat[\centering Estimated $\widehat{\beta}_\text{NST-GAM}$ (MISE: $19.91$)]{{\includegraphics[height = 1.6in, width=2.9in]{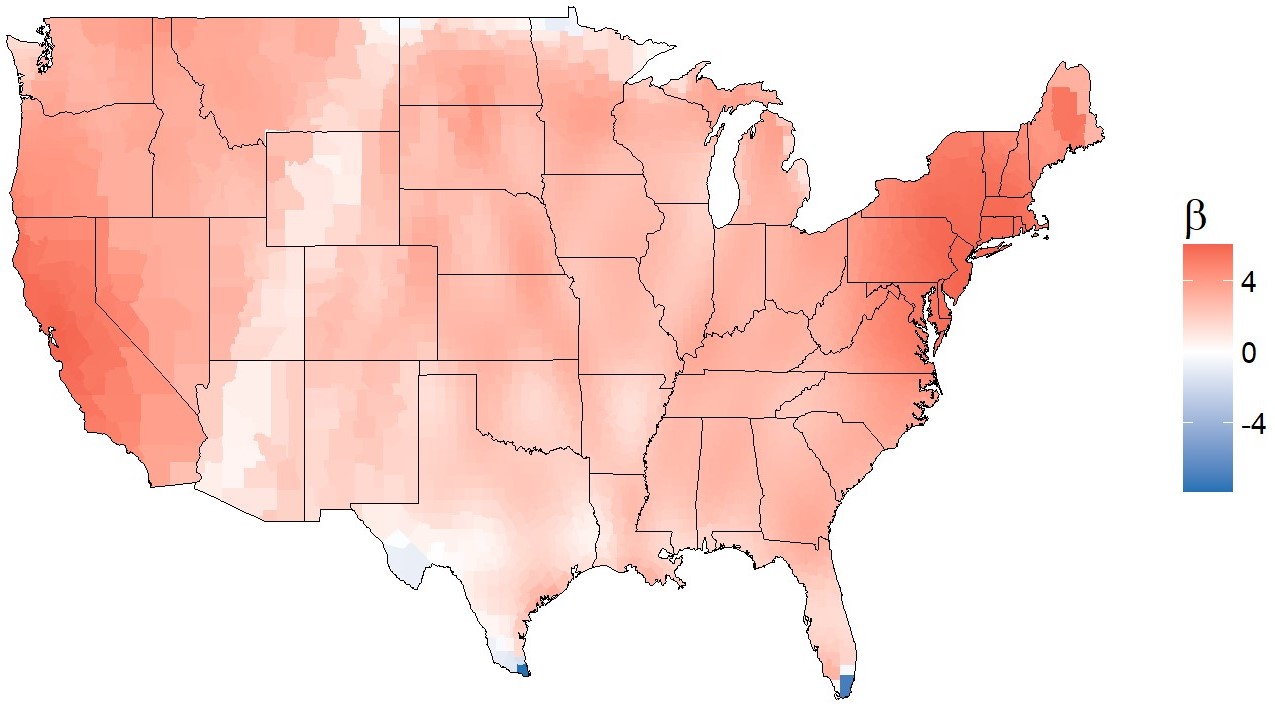} }}%
    \caption{True and estimated bivariate functions comparison in numerical study 2}%
    \label{fig:Covid_Experiement_Bivariate}%
\end{figure}

\begin{figure}[H]
\captionsetup{font=small}
\begin{center}
\begin{tabular}{cc}
\includegraphics[height=3.3in, width = 6.235in]{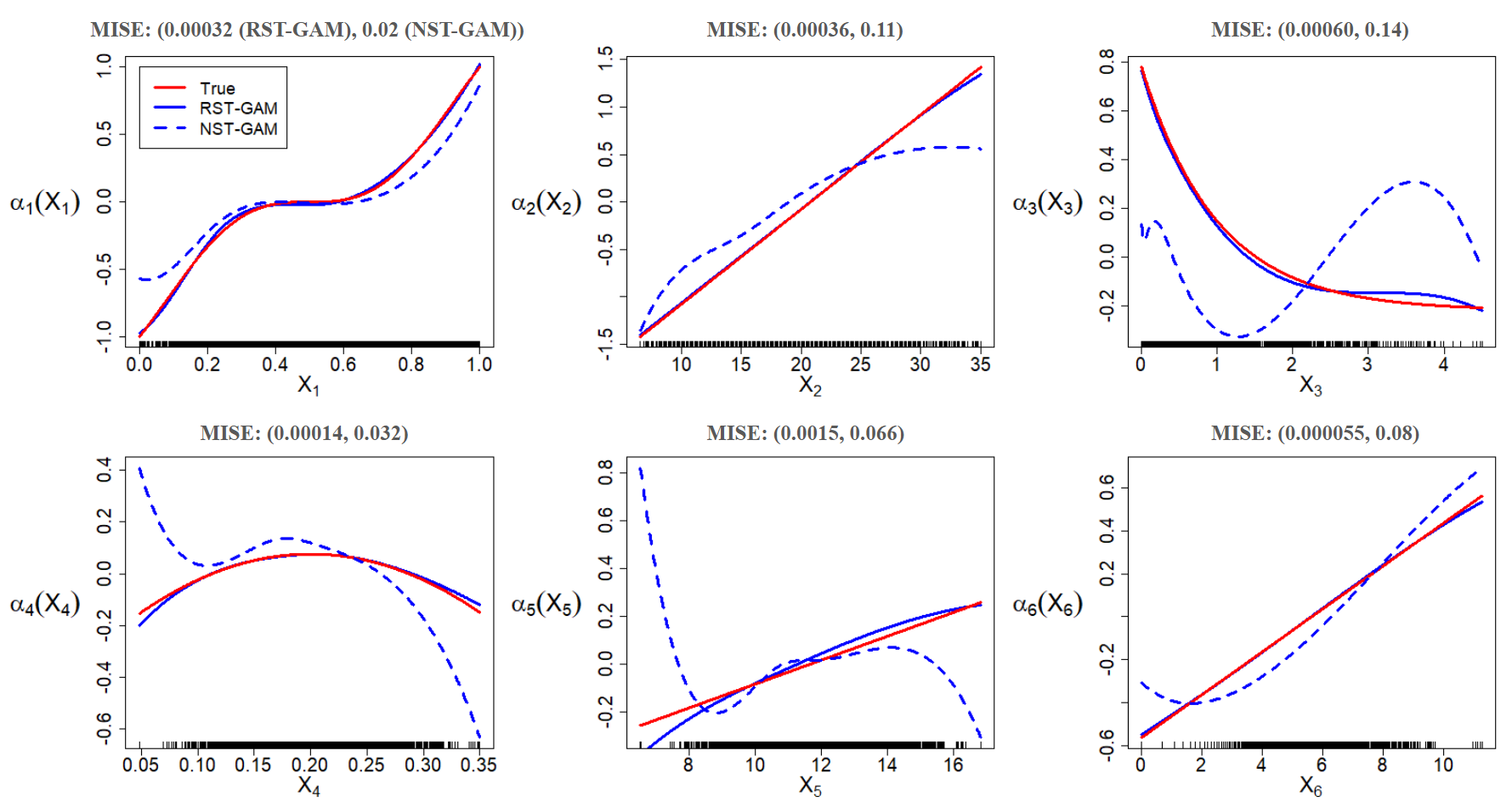}
\end{tabular}
\end{center} 
\caption{True and estimated univariate components comparison in numerical study 2} 
\label{fig:Covid_Experiment_Univariate}
\end{figure}

To design an experiment that realistically mimics a pandemic situation with a complex spatial domain, we utilize a subset of covariates and the domain from Section \ref{sec:Real_Covid} to create a synthetic county-level infection counts dataset. The response $Y_{it}$ is generated from a Poisson distribution as $Y_{it} \sim \text{Poisson}(\mu_{it})$, where $\mu_{it} = \exp\left(\beta_{t}(\V{u}_i) + \sum_{k=1}^{6}\alpha_{kt}(X_{itk})\right)$ and $\{X_{itk}\}_{k = 1}^{6}$ are selected from the COVID-19 datasets described in Table~\ref{table:covid_covariates} for $i = [3104]$ and $\,t = [5]$. Figure~\ref{fig:Covid_Experiement_Bivariate} (a) shows the true bivariate component $\beta_t$. Figure~\ref{fig:Covid_Experiment_Univariate} shows the true
univariate components $\{\alpha_{kt}\}_{k=1}^6$ (red curves). To introduce outliers, we randomly select 200 counties across the spatial domain and add a constant shift of 200 to their corresponding mean parameter $\mu_{it}$ over the entire observation period.  Figure~\ref{fig:Covid_Experiement_Bivariate} (b) displays the spatial distributions of the simulated outliers.

\vspace{2mm}

Table~\ref{table:Outlier_Detection_for_Simulated_COVID_Data} summarizes the outlier detection results. \texttt{RST-GAM} successfully identifies all outliers in the data with only a single misclassification of a non-outlier point. We also evaluate the accuracy of the estimated functional forms. Figure~\ref{fig:Covid_Experiement_Bivariate} (c) and (d) demonstrate that \texttt{RST-GAM} outperforms the baseline method, estimating the spatial transmission pattern more accurately. Similar patterns are observed in the univariate component estimates.  Figure~\ref{fig:Covid_Experiment_Univariate} shows that \texttt{NST-GAM} estimates near-zero effects for some covariates, failing to capture the true functional forms in these covariates. These results suggest that even a small proportion of outliers (6\% in this case) can substantially bias the estimated component functions. By contrast, \texttt{RST-GAM} can recover those underlying signals with high precision despite the outliers.

\vspace{1.5mm}
\begin{table}[H]
\captionsetup{font=small}
\centering
\setlength{\tabcolsep}{10pt} 
\renewcommand{\arraystretch}{1.2} 
\caption{Outlier Detection in simulated infections data}%
\begin{tabular}{clcc}
\multicolumn{1}{l}{}     & \multicolumn{3}{c}{True}                                                    \\ \hline
\multirow{3}{*}{Predicted} &             & \multicolumn{1}{l}{Outlier} & \multicolumn{1}{l}{Non-outlier} \\ \cline{2-4} 
                         & Outlier     & 200                          & 1                               \\ \cline{2-4} 
                         & Non-outlier & 0                           & 2903                            \\ \hline
\end{tabular}
\label{table:Outlier_Detection_for_Simulated_COVID_Data}
\end{table}

\section{Application to COVID-19 Data}
\label{sec:Real_Covid}
During the COVID-19 pandemic, county-level daily infection counts were collected to monitor the dynamics of the disease spread. For the most part, these infection counts exhibit a relatively smooth spatial variation across locations, as neighboring counties tended to share similar spatial and covariates' effects. This smooth pattern was not always observed however, as some counties reported significantly higher infection counts than their neighbors. This was especially true in the early stages of the pandemic. Such spikes, even if infrequent, can impact the identification of transmission patterns and our understanding of covariate effects on infection counts.

We have two goals in this study. First, we want to detect the locations for possible outlying counties. This is especially important at the beginning of a pandemic. Early identification can inform resource allocations as those counties could be critical to controlling the spread of disease. At the same time, we want to identify the spatial distribution of new infections and examine how different socioeconomic, healthcare, demographic, and epidemiological factors contribute to the COVID-19 transmission after adjusting for the influences of these outliers.
\vspace{3mm}

\begin{table}[H]
\captionsetup{font=small}
\centering
\setlength{\tabcolsep}{8pt} 
\renewcommand{\arraystretch}{1.2} 
\caption{County-Level covariates in the COVID data analysis}
\small
\begin{tabular}{llp{7.4cm}}
\hline
\textbf{Variable Group} & \textbf{Variable Name} & \textbf{Description} \\ \hline
\multirow{4}{*}{\centering Socioeconomic Status} & Affluence & Social affluence measure \\
                                                & Gini & Gini index\\
                                                & Disadvantage & Economic disadvantage measure \\
                                                & RUrban & Urbanization rate \\ \hline
\multirow{3}{*}{\centering Healthcare Resources} & ExpHealth & Goverment expenditure on healthcare per capita \\
                                                  & RHealCov & Percentage of population without health insurance \\
                                                  & Bed & Total number of beds per 1000 population (in logarithmic scale)\\ \hline
\multirow{5}{*}{\centering Demographic Information} & RAA  & Proportion of African American in population\\
                                                    & RHL  & Proportion of Hispanics or Latinx in population \\
                                                    & RMF & Male to female ratio \\
                                                    & Pop & Population per square mile (in logarithmic scale) \\
                                                    & ROld & Proportion of population with age $\geq 65$ \\ \hline
\multirow{2}{*}{\centering Epidemiological Dynamics} & Mobility & Daily number of trips within
each county in previous week (in logarithmic scale)\\
                                                    & AccInfected & Accumulated inferctions on previous day (in logarithmic scale) \\ \hline
\end{tabular}
\label{table:covid_covariates}
\end{table}
We utilized daily infection COVID-19 counts at the county-level over 3,104 counties across the mainland U.S.\@ from \cite{wang2022} for this study. We excluded regions outside the contiguous United States, such as Alaska and Hawaii, due to their geographic separation. The daily infection data for these counties were sourced and cleaned from the state health departments, the Center for Systems Science and Engineering at Johns Hopkins University, and the COVID Tracking Project. Spatial information and 14 county-level covariates were collected from the U.S. Census Bureau and the U.S. Department of Homeland Security. 
Table~\ref{table:covid_covariates} provides a detailed summary of these features.

\begin{figure}[H]
\captionsetup{font=small}
\begin{center}
\begin{tabular}{cc}
\includegraphics[height=1.5in, width = 2.5in]{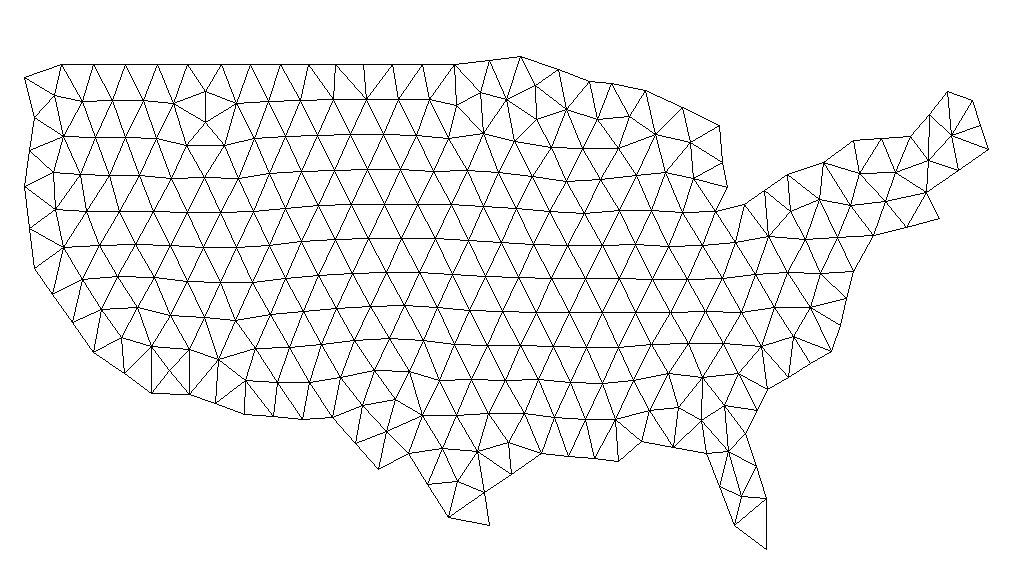}
\end{tabular}
\end{center} 
\caption{Triangulation of mainland U.S.} 
\label{fig:covid_triangulation}
\end{figure}

To minimize the impact of different COVID-19 variants on our analysis, we focus exclusively on the period when the wild-type strain was predominant by using data from April 1, 2020 to April 30, 2021 \citep{CDC_COVID_Timeline}. For a given location $i$ and time $t$, we consider the following model
\begin{align*}
\log(\mu_{it}) \amp = \amp & \beta_{t}(\V{u}_i) + \alpha_{1t}(\text{Affluence}_{i}) + \alpha_{2t}(\text{Gini}_{i}) + \alpha_{3t}(\text{Disadvantage}_{i}) + \alpha_{4t}(\text{RUrban}_{i})  \\  
& + \alpha_{5t}(\text{ExpHealth}_{i}) + \alpha_{6t}(\text{RHealCov}_{i}) + \alpha_{7t}(\text{Bed}_{i}) \\
& + \alpha_{8t}(\text{RAA}_{i}) + \alpha_{9t}(\text{RHL}_{i}) + \alpha_{10t}(\text{RMF}_{i}) + \alpha_{11t}(\text{Pop}_{i}) + \alpha_{12t}(\text{ROld}_{i}) \\
& + \alpha_{13t}(\text{Mobility}_{i,t-7}) + \alpha_{14t}(\text{AccInfected}_{i, t-1}), 
\end{align*}
where $\{\alpha_{kt}\}_{k=1}^{14}$ are univariate functions that model the association of daily infections with the covariates in Table \ref{table:covid_covariates}. For our \texttt{RST-GAM}, we applied a 3-fold data-thinning approach to construct the weights for the adaptive Lasso. Figure~\ref{fig:covid_triangulation} shows the triangulation of the mainland U.S. that we used. It consists of 522 triangles and 306 vertices. To estimate the univariate functions, we used a cubic spline with 2 interior nodes. We used the same triangulation and cubic spline setup for \texttt{NST-GAM}.

We use a 7-day observation window to analyze the dynamics of disease spread and the covariates' effects on the daily reported infection cases. According to the Centers for Disease Control and Prevention (CDC), the prevalence of the wild-type strain of COVID-19 can be categorized into three key periods: the early stage of the pandemic (starting in early April 2020), the peak phase (beginning in early November 2020), and the stabilization phase, marked by a decline in cases after March 2021 \citep{CDC_COVID_Timeline}.

\vspace{1.5mm}
\begin{figure}[H]
\captionsetup{font=small}
    \centering
    \subfloat[\centering Early phase - Number of identified outlying counties: 31]{{%
    \includegraphics[height = 1.6in, width=3in]{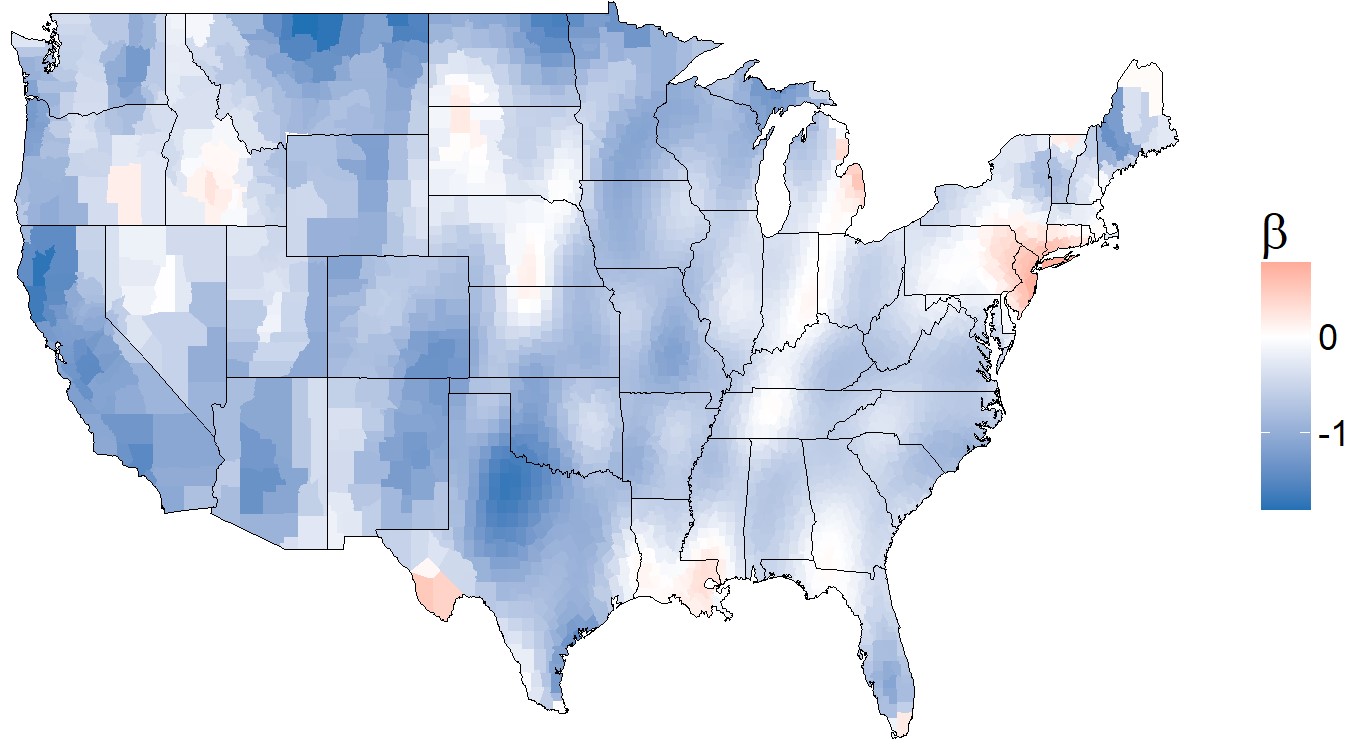}%
    \hspace{0.7cm} 
    \includegraphics[height = 1.6in, width=3in]{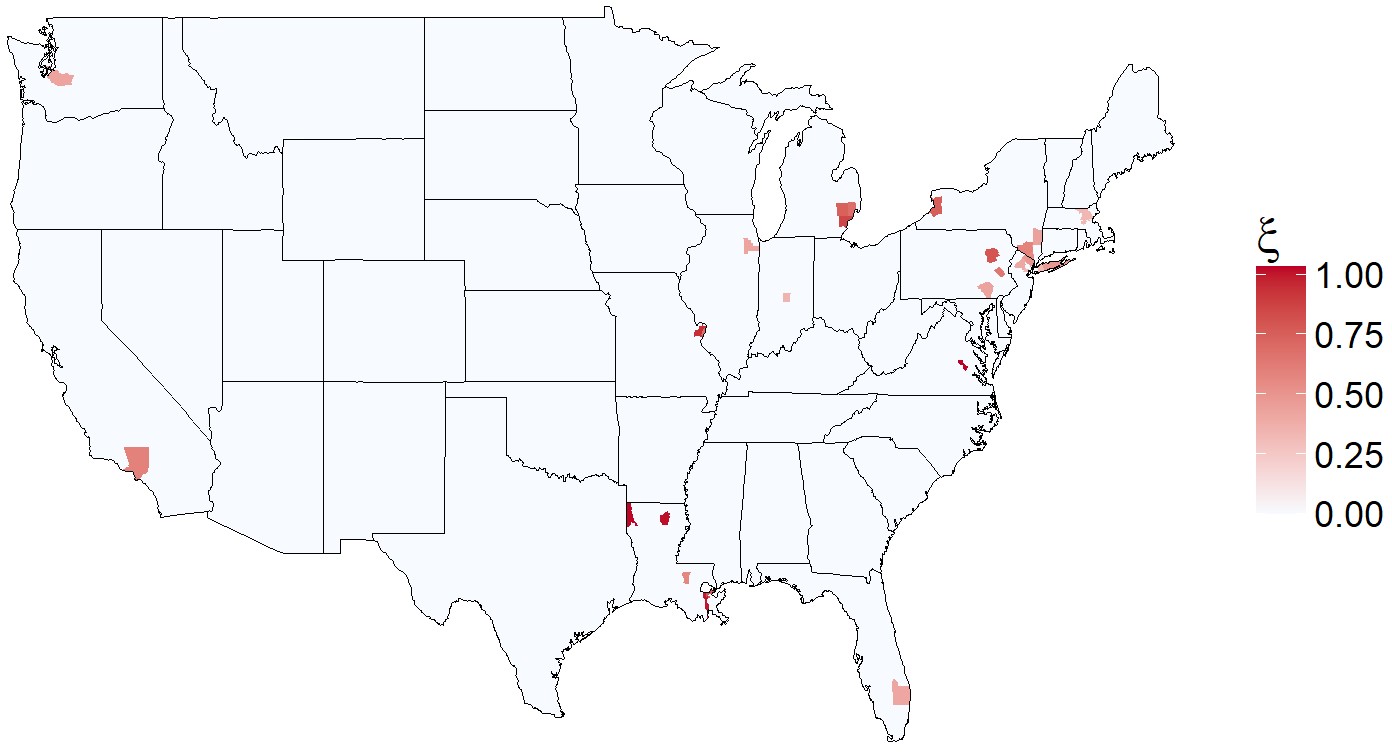} }}%
    \vspace{0.7cm}
    \subfloat[\centering Peak phase - Number of identified outlying counties: 618]{{%
    \includegraphics[height = 1.6in, width=3in]{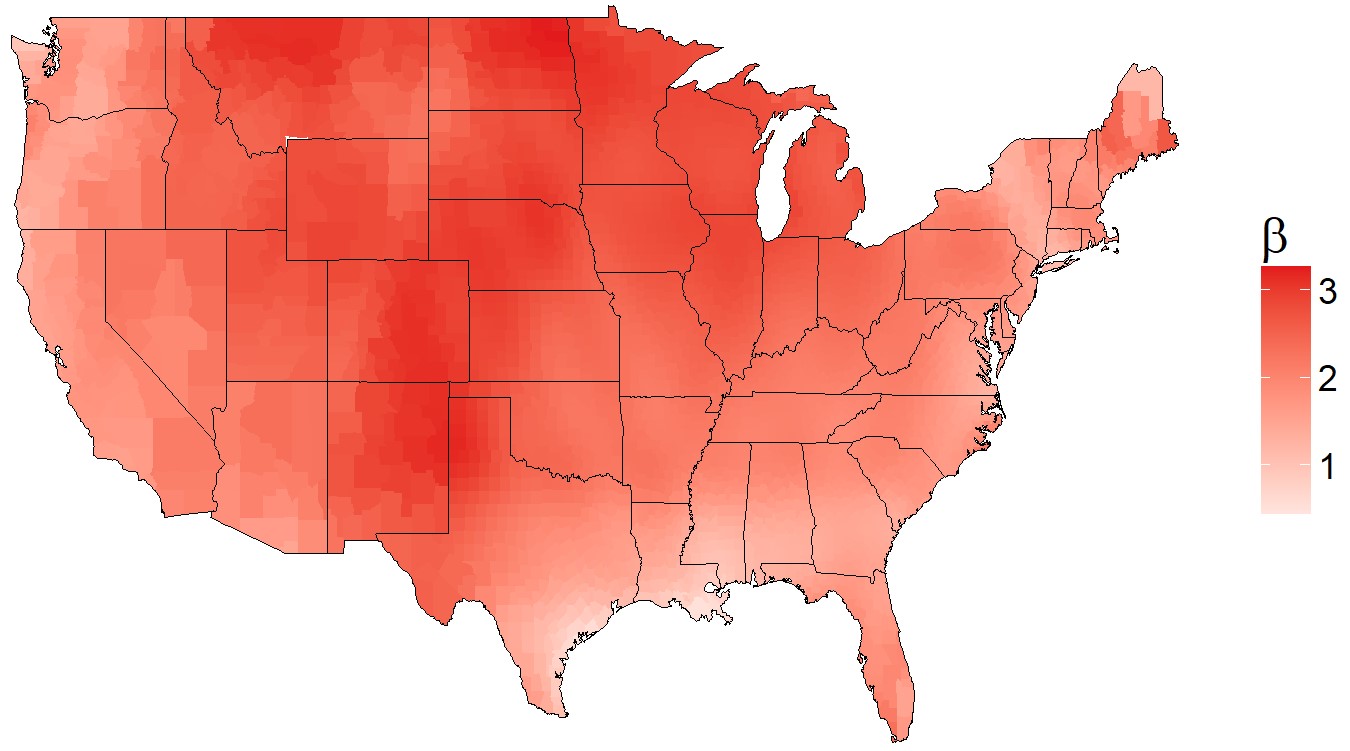}%
    \hspace{0.7cm} 
    \includegraphics[height = 1.6in, width=3in]{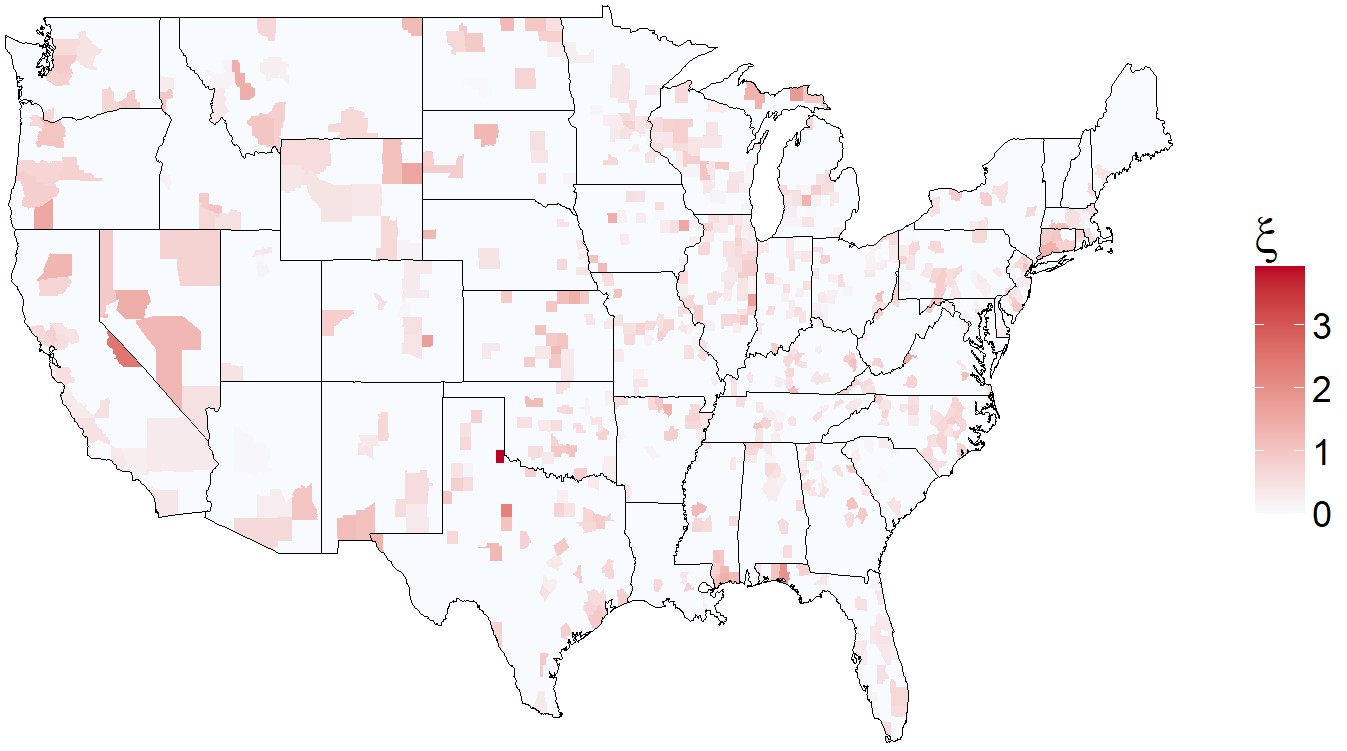} }}%
    \vspace{0.7cm}
    \subfloat[\centering Stabilization phase - Number of identified outlying counties: 451]{{%
    \includegraphics[height = 1.6in, width=3in]{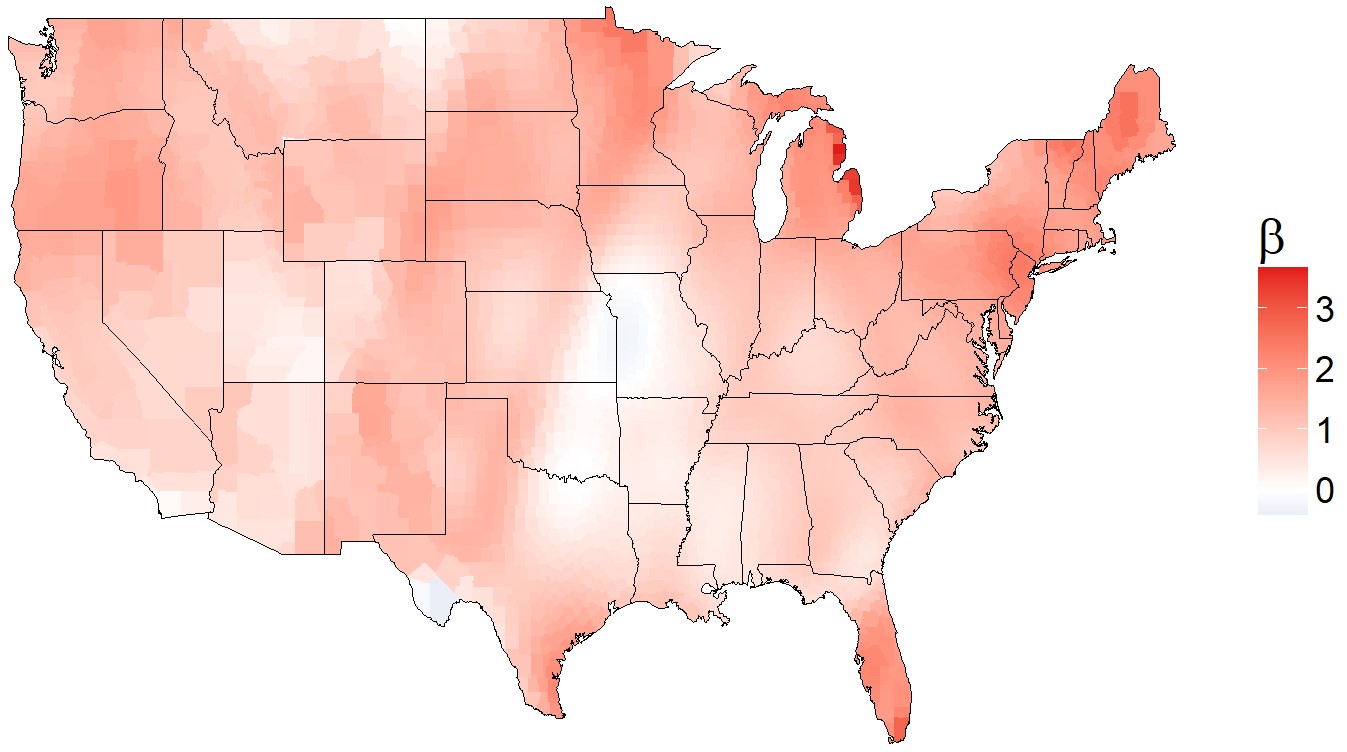}%
    \hspace{0.7cm} 
    \includegraphics[height = 1.6in, width=3in]{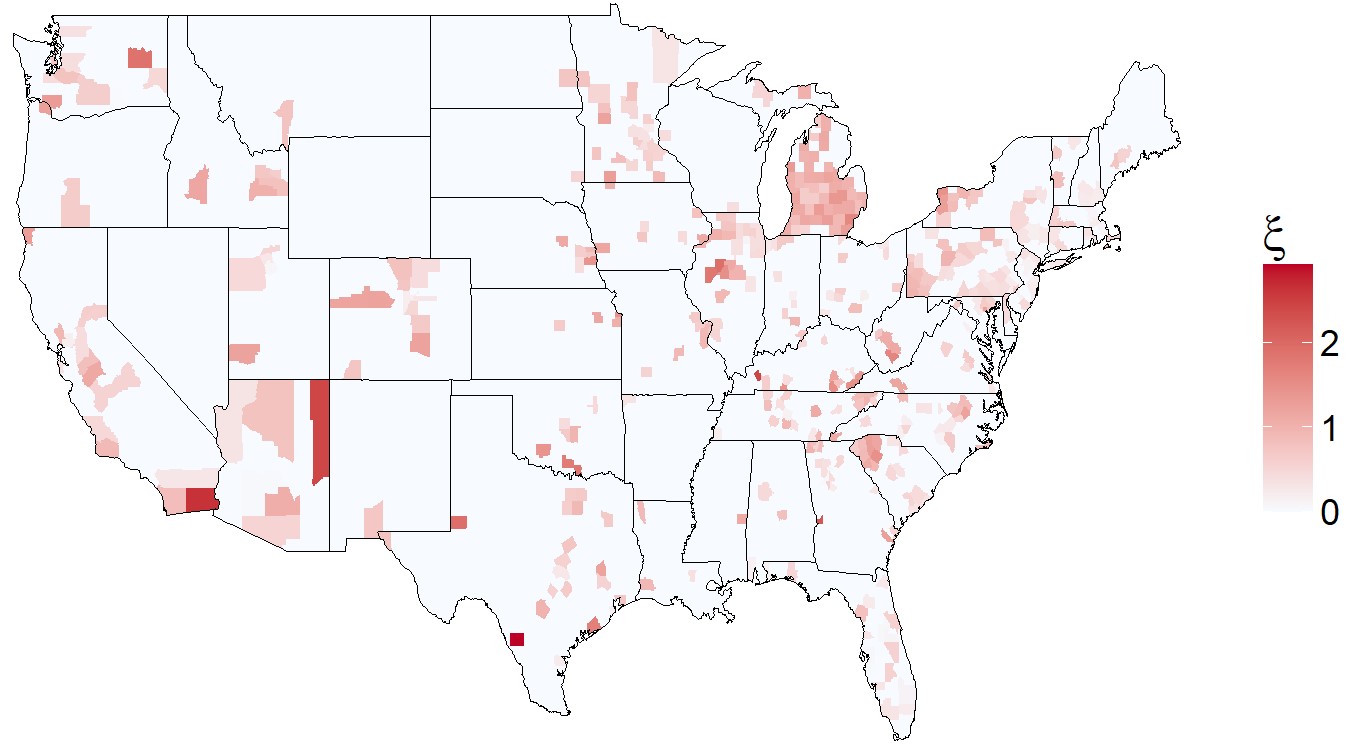} }}%
    \vspace{0.3cm}
    \caption{\texttt{RST-GAM}'s estimates of bivariate functions (left) and identified outlying counties (right) over 3 key periods categorized by the CDC}%
    \label{fig:Covid_Analysis_Bivariate}%
\end{figure}

Figure~\ref{fig:Covid_Analysis_Bivariate} displays a heatmap of the estimated spatial components and identified outliers for one time slice within each of these three periods. During the early stage of the pandemic, \texttt{RST-GAM} reveals positive spatial effects on new infections, concentrated in the New York-Pennsylvania region and the New Orleans area. Meanwhile, outliers are predominantly identified in coastal counties, consistent with the early spread of COVID-19 being concentrated along the U.S. coastline. Interestingly, \texttt{RST-GAM} detects a few inland outliers, such as St. Louis County in Missouri, suggesting potential inland hotspots with substantially higher daily infections compared to neighboring areas. As the pandemic progressed to its peak in November 2020, \texttt{RST-GAM} identifies widespread positive spatial effects across most states, accompanied by a dramatic increase in outliers nationwide. This observation aligns with the pandemic's peak, characterized by surging infections across the country. By April 2021, the estimated spatial function shows a notable decline across all states, along with a reduction in the number of detected outliers, reflecting the stabilization and gradual decline of the pandemic.

\begin{figure}[H]
\captionsetup{font=small}
\begin{center}
\begin{tabular}{cc}
\includegraphics[height=3.8in, width = 6.3in]{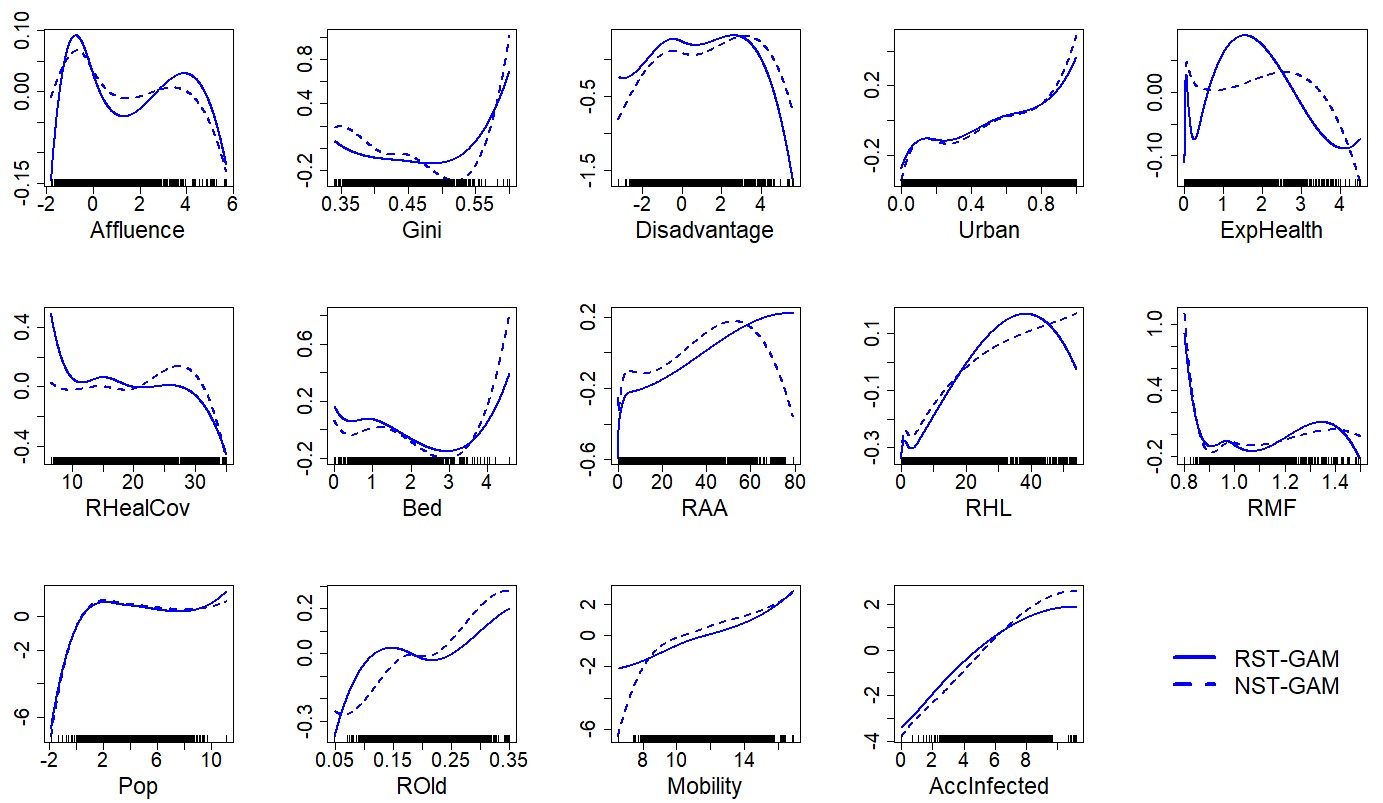}
\end{tabular}
\end{center} 
\caption{Estimated univariate functions in the early phase of COVID-19 pandemic (4/1/2020 - 4/7/2020)} 
\label{fig:real_Covid_beginning_Univariate}
\end{figure}

Next, we examine the effects of the 14 covariates on the spread of COVID-19. Figure~\ref{fig:real_Covid_beginning_Univariate} presents the estimated univariate functions during the early stage of the COVID-19 pandemic. The mobility covariate demonstrates a strong positive association with newly infected cases, as high mobility levels are often associated with increased interactions and opportunities for viral spread \citep{Kraemer2020Mobility}. A similar positive trend is observed for accumulated infection counts, which is expected since areas with high initial infections tend to experience sustained outbreaks due to community transmission chains \citep{deng2020genomic}. The proportions of African American and Hispanic or Latinx populations also exhibit significant positive associations, which reflects the disproportionate burden of COVID-19 on minority communities \citep{magesh2021disparities}. On the other hand, no strong signals are observed for affluence and level of economic disadvantage, as studies point out that early infection dynamics were influenced more by spatial and demographic factors than by socioeconomic disparities \citep{jackson2021spatial}. The male to female ratio has an estimated effect near zero, suggesting that infection risk is similar across genders.

Notably, \texttt{RST-GAM} identifies a significant negative effect for the percentage of the uninsured population. Such association could be attributed to the situation of underreporting.
Uninsured individuals may face financial or logistical barriers to accessing COVID-19 testing or healthcare services, leading to fewer diagnosed and reported cases \citep{albani2021covid}. By contrast, the \texttt{NST-GAM} model estimates a negligible effect for this variable. Such results align with conclusions from the numerical studies in Section~\ref{subsec:exp_2}); outliers can obscure the identification of univariate relationships. This discrepancy suggests that \texttt{RST-GAM} can enable a more nuanced understanding of the spread of disease by adjusting for the influence of outliers.

\begin{figure}[H]
\captionsetup{font=small}
    \centering
    \subfloat[\centering Urbanization Rate]{{\includegraphics[height=1.6in, width = 6in]{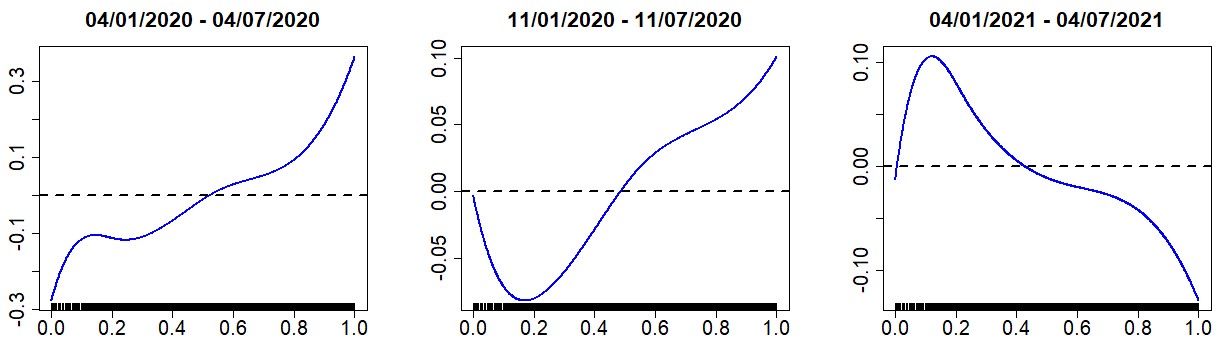} }}%
    \vspace{0.3cm} 
    \subfloat[\centering Percentage of Aged Population]{{\includegraphics[height=1.6in, width = 6in]{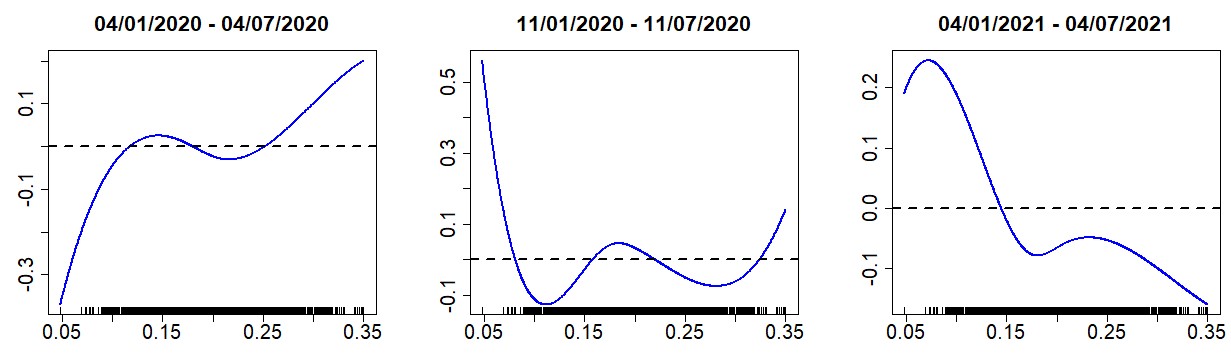} }}%
    \qquad    
    \caption{\texttt{RST-GAM}'s estimates of univariate functions over 3 key periods categorized by the CDC}%
    \label{fig:Covid_time_tread_univariate}%
\end{figure}

Finally, we examine the contributions of covariates to COVID-19 spread over time, focusing on urbanization rate and the proportion of aged populations (Figure~\ref{fig:Covid_time_tread_univariate}). In the early phase of the pandemic, urban areas played a critical role as transmission hubs due to increased human interactions, reflected in a strong positive effect of urbanization. This effect persisted through the peak phase but weakened substantially during the stabilization phase, likely due to vaccination efforts and public health measures reducing transmission. Similarly, the proportion of aged populations had a moderate positive effect during the early phase, indicating heightened vulnerability among seniors. This effect, however, nearly vanished during the peak phase, possibly due to behavioral changes like reduced mobility and adherence to protective measures. By the stabilization phase, seniors emerged as the least affected group, highlighting the success of targeted vaccination campaigns in reducing their susceptibility \citep{Dooling2021CDC}.

\vspace{1.5mm}

\section{Discussion}
\label{sec:conclusion}
In this work, we introduced \texttt{RST-GAM}, a robust statistical framework for epidemic modeling. \texttt{RST-GAM} achieves robustness by adding sparse slack variables to a spatiotemporal GAM. Imposing an adaptive Lasso penalty provably enhances precision in outlier identification, and our novel use of data-thinning techniques enables a practical and effective construction of data adaptive weights. We provide a provably convergent variant of PGD for efficiently estimating the \texttt{RST-GAM} parameters. Our numerical studies demonstrate that \texttt{RST-GAM} can achieve robust estimation and accurate outlier detection simultaneously. Our analysis of COVID-19 data also demonstrates \texttt{RST-GAM}'s capacity to mitigate outlier-driven distortions and provide more precise insights into the factors influencing COVID-19 spread. By contrast, the baseline approach fails to capture these nuances. 

We close with remarks on ways to build on \texttt{RST-GAM}'s capabilities. First, developing a multivariate version of \texttt{RST-GAM} would facilitate the simultaneous exploration of multiple COVID-19 variants, such as Delta and Omicron, and enable a complete picture of transmission behaviors when multiple variants are circulating. Second, introducing epidemic compartments like susceptible, exposed, and recovered groups would enable a better understanding of disease dynamics and consequently provide the basis for improved interventions \citep{Brauer2008, wang2022}. Finally, deriving simultaneous confidence bands \citep{mckeague2006width, wang2009polynomial}, would enable rigorous inference on the estimators. We leave these extensions for future work.

An R-package `RSTGAM` implementing the methods described in this article
is available at \url{https://github.com/haomingsj98/RSTGAM}.

\section{Funding}
Haoming Shi is partially supported through the AI2Health research cluster of the Ken Kennedy Institute at Rice University.
\newpage

\bibliographystyle{apalike}
\bibliography{refs}

\end{document}